\documentclass[journal]{IEEEtran}
\usepackage{cite}
\usepackage[pdftex]{graphicx}
\usepackage{amssymb,amsmath}
\usepackage{float}
\usepackage{bm}
\usepackage{mathtools}
\usepackage{amsmath}
\usepackage{array}
\usepackage{multirow}
\usepackage{stfloats}
\usepackage[utf8]{inputenc}
\usepackage[english]{babel}
\usepackage{array}
\usepackage{mathrsfs}
\usepackage{makecell}
\usepackage{tabularx,booktabs}

\usepackage[most]{tcolorbox}
\usepackage{color, colortbl}
\usepackage{pifont}

\usepackage{algorithm,algpseudocode}
\makeatletter
\renewcommand{\Function}[2]{%
  \csname ALG@cmd@\ALG@L @Function\endcsname{#1}{#2}%
  \def\jayden@currentfunction{#1}%
}
\newcommand{\funclabel}[1]{%
  \@bsphack
  \protected@write\@auxout{}{%
    \string\newlabel{#1}{{\jayden@currentfunction}{\thepage}}%
  }%
  \@esphack
}
\let\myorg@bibitem\bibitem
\def\bibitem#1#2\par{%
  \@ifundefined{bibitem@#1}{%
    \myorg@bibitem{#1}#2\par
  }{%
    \begingroup
      \color{\csname bibitem@#1\endcsname}%
      \myorg@bibitem{#1}#2\par
    \endgroup
  }%
}
\makeatother

\begin{document}
\title{Unified Virtual Oscillator Control for Grid-Forming and Grid-Following Converters}

\author{M A~Awal,~\IEEEmembership{Student Member,~IEEE,}
        and~Iqbal~Husain,~\IEEEmembership{Fellow,~IEEE}

FREEDM Systems Center, North Carolina State University, Raleigh, NC 27695, USA
\thanks{This work has been supported in part by the National Science Foundation under award number EEC-0812121 for the FREEDM Engineering Research Center.}

\thanks{The authors are with the FREEDM Systems Center, North Carolina State University, Raleigh, NC 27695 USA (e-mail:, mawal@ncsu.edu; ihusain2@ncsu.edu).}
}

\markboth{Awal et. al.: Unified Virtual Oscillator Control for Grid-Forming and Grid-Following Converters}
{}

\maketitle

\begin{abstract}
A unified virtual oscillator controller (uVOC) is proposed, which enables a unified analysis, design, and implementation framework for both grid-forming (GFM) and grid-following (GFL) voltage source converters (VSCs). Oscillator based GFM controllers, such as dispatchable virtual oscillator control (dVOC), offer rigorous analytical framework with enhanced synchronization, but lack effective fault handling capability which severely limits practical application. The proposed uVOC facilitates synchronization with an arbitrarily low grid voltage and fast over-current limiting; this enables effective fault ride-through unlike existing GFM controllers which typically switch to a back-up controller during fault. GFM operation with uVOC is achieved in both grid connected and islanded modes with seamless transition between the two.  In GFL converters, bidirectional power flow control and DC bus voltage regulation is achieved with uVOC. No phase-locked-loop (PLL) is required for either GFL or GFM operation circumventing the synchronization issues associated with PLLs in weak grid applications. Detail small signal models for GFM and GFL operation have been developed and systematic design guidelines for controller parameters are provided. The proposed controller is validated through hardware experiments in a hybrid AC-DC microgrid.

\end{abstract}

\vspace{10pt}

\begin{IEEEkeywords}
unified virtual oscillator control, uVOC, oscillator based control, grid forming converter, grid following converter, weak grid, fault ride-through
\end{IEEEkeywords}

\IEEEpeerreviewmaketitle

\bstctlcite{IEEEexample:BSTcontrol}

\section{Introduction}
\label{sec:introduction}
With the increasing penetration of distributed generation resources into the power system, coordinated control of low-inertia systems has garnered significant research efforts\cite{LIN1, LIN2, LIN3}. Grid-forming (GFM) converters (GFMC) are considered to be a key enabling technology for the future grid with reduced inertia \cite{GFMC}. The prevalent and most widely investigated control strategies for GFMCs have been developed based on mimicking the characteristics of synchronous machines by voltage source converters (VSCs). Droop control\cite{dc0,dc1}, power synchronization control (PSC) \cite{psc_org}, synchronverter\cite{VSM1}, virtual synchronous machine (VSM/VISMA)\cite{VISMA1}, and synchronous power controller (SPC)\cite{spc1} are examples of emulation based GFM control methods. Such machine emulation based methods are beneficial since they facilitate intuitive converter level design and implementation. However, it is yet to be proven and demonstrated whether machine emulation based methods provide the best solution for GFMC control in low-inertia networks. 
To circumvent the phasor based approximate modelling and design, a class of nonlinear time-domain GFM control methods, offering rigorous analytical framework, have been recently proposed\cite{VOC1,VOC2,VOC3,VOC4,TPELVoc,dVOC1,dVOC2,dVOC3,dVOC4}. The so-called dispatchable virtual oscillator control (dVOC), inspired by consensus in multi-agent networks, offers almost global synchronization guarantee in a network with an arbitrarily large number of converters and hence is a promising research direction for GFMC control in future grids with ultra low inertia. 
However, to the best of the authors' knowledge, no compatible fault ride-through strategy has been reported for oscillator based GFM controllers, which severely limits practical application. Moreover, fault-handling in GFM converters, even in conventional droop based VSCs, remains an open research problem till date. In a PSC based converter, a backup PLL is run for fault management; during grid faults the converter control system switches from PSC to PLL based current controlled operation\cite{psc_org}. In droop based GFM converters, over-current protection may be achieved by directly limiting the reference to the inner current control loop; but this causes wind-up in the outer droop control loops which eventually leads to loss of synchronization and instability \cite{FH1}. As an alternative, switching to PLL based grid-following operation during faults has been proposed in \cite{FH2,FH3}; dynamic virtual impedance control and adaptive droop control during faults were proposed in \cite{FHWang}.
\par For effective fault ride-through, two key capabilities are required - first, synchronization with arbitrarily low grid voltage and second, fast over-current limiting (OCL). In this work, we build on the rigorous synchronization results of dVOC and propose an oscillator which enables synchronization with grid voltage of arbitrary magnitude. The ensuing controller is labeled as unified virtual oscillator controller (uVOC). In addition to fault ride-through, this form of the oscillator enables grid-following operation without a PLL. Sub-synchronous oscillations or even instability triggered by PLLs in weak grid conditions \cite{PLLuwg,psc_org} are well known which can be avoided leveraging the proposed controller. Fast OCL capability is achieved by introducing a series compensator cascaded with the oscillator. We develop a unified analysis, control design, and implementation framework for GFM and GFL operation which enables fault ride-through without the need for switching to a back-up controller. Furthermore, PLL-less grid synchronization in GFL converters with bi-directional power flow control and DC bus voltage regulation is achieved.

It is worth noting that oscillator based time-domain controllers are a class of emerging technology. Performance comparison of oscillator based time-domain controllers with machine emulation based methods can be found in \cite{vocVsDc1,vocVsDc2}, where significant distinctions in dynamic response are observed between the two despite identical steady-state droop settings. In this work, we develop the analysis and design framework for uVOC building up from the theoretical results for GFM operation developed in \cite{dVOC1,dVOC2,dVOC3, dVOC4}; performance comparison with phasor-domain methods require further comprehensive study and is not within the scope of this work.   

\par The rest of the paper is organized as follows: first, a space vector based nonlinear oscillator is described for grid synchronization in both GFL and GFM operation. Second, the overall implementation structure of uVOC including fault management, emulated virtual impedance and pre-synchronization method is introduced. Third, through steady state and small signal analysis, controller design guidelines are presented. Lastly, simulation and experimental results are presented to validate the proposed controller.

\begin{figure}[htb]
	\makebox[\linewidth][c]{\includegraphics[angle = 0, clip, trim=0cm 0cm 0cm 0cm,  width=0.4\textwidth]{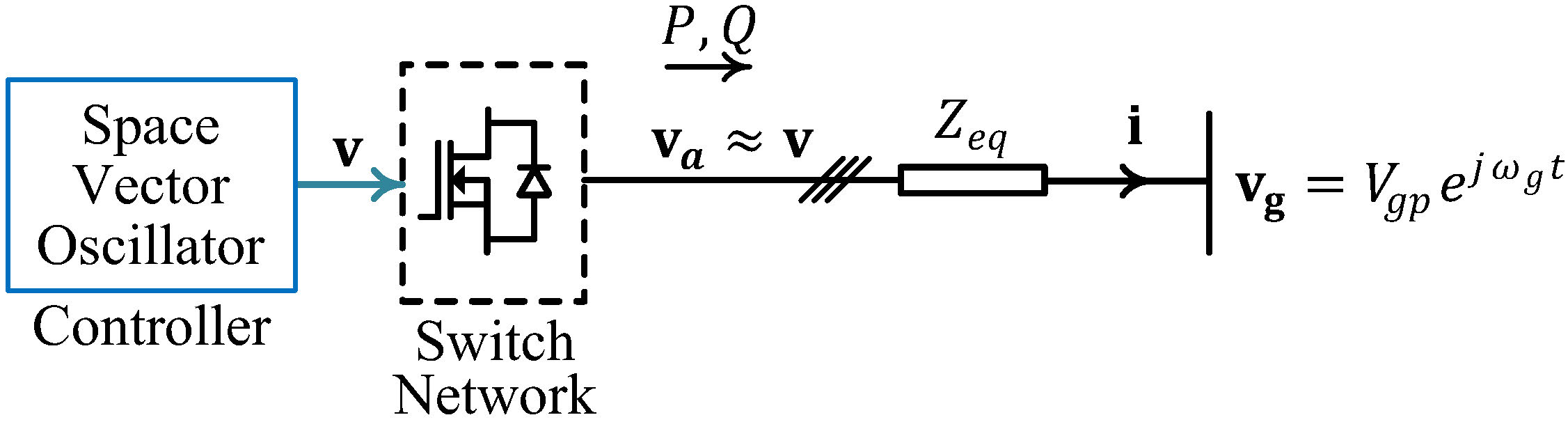}}
	\caption{Simplified VSC system connected to an infinite bus through an equivalent impedance for controller development.}
	\label{fig:simSysDes}
\end{figure}

\section{Space Vector Oscillator (SVO)}\label{sec:svo}
The SVO serves as the synchronization unit for uVOC. For the purpose of development of the SVO, we consider a simplified VSC system, shown in Fig.~\ref{fig:simSysDes}, connected to an infinite bus $\mathbf{v_g}=V_{gp}e^{j\omega_g t}$ through an equivalent impedance $Z_{eq}$; the controller implementation in a real VSC is presented in Section~\ref{sec:ImpIssues}. Taking average over a switching period, the switch network can be approximated as a unity-gain block, i.e., $\mathbf{v_a}\approx \mathbf{v}$; here, the three-phase voltage $[v_U\ v_V\ v_W]^T$ is represented as a space vector $\mathbf{v}=v_\alpha + jv_\beta \leftrightarrow [v_\alpha\ v_\beta]^T$ in the stationary $\alpha \beta$-reference frame, where $j=\sqrt{-1}$ denotes the imaginary unit. Similar space vector notation is followed in the discussion that follows. 
The problem statement can be defined as to determine a dynamic control law on the converter output voltage vector $\mathbf{v}$ which enables synchronization in GFL and GFM converters.

\begin{figure}[htb]
	\makebox[\linewidth][c]{\includegraphics[angle = 0, clip, trim=0cm 0cm 0cm 0cm,  width=0.275\textwidth]{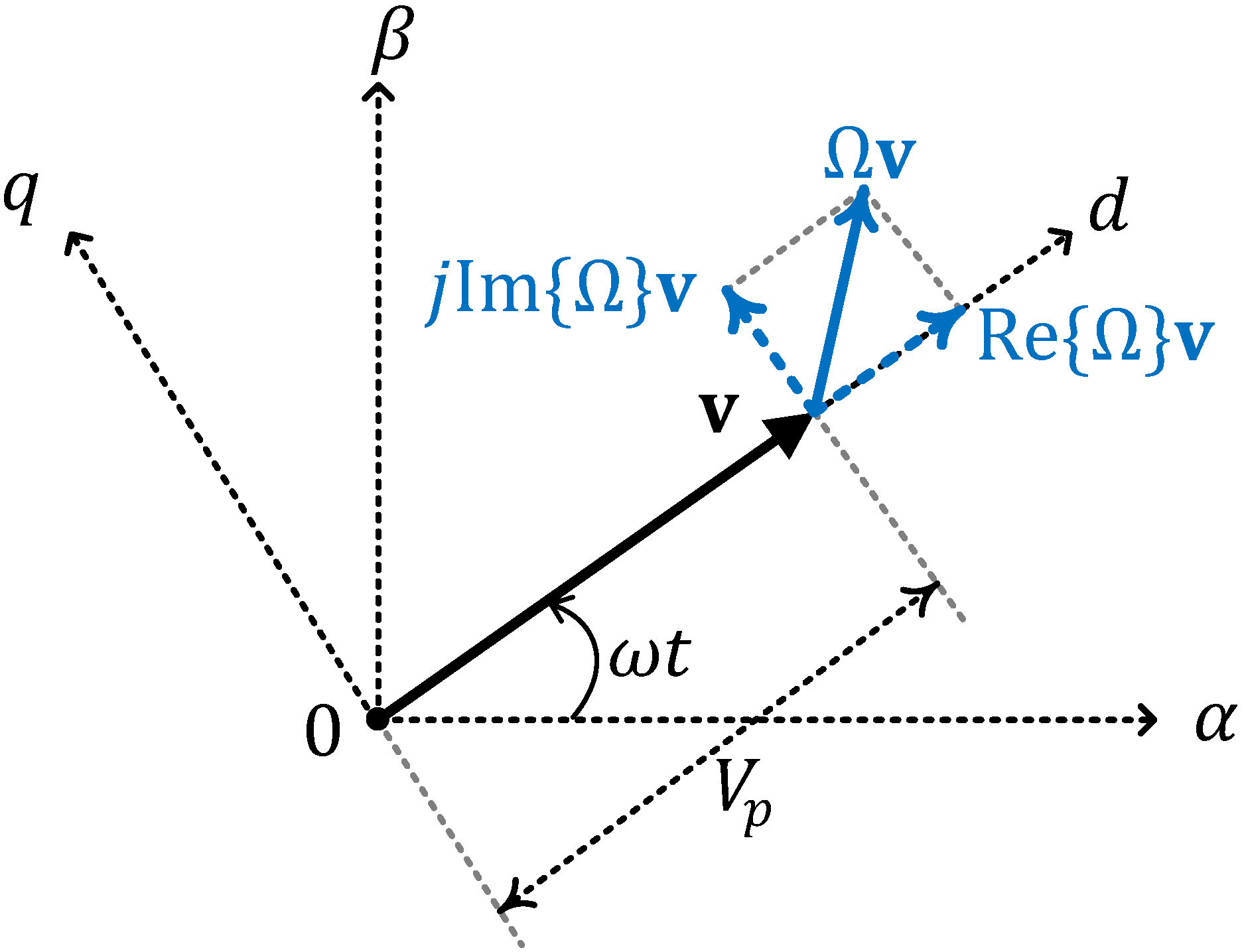}}
	\caption{The real and imaginary parts of the time-derivative are exerted along the $d$ and $q$ axes on the original space-vector, respectively.}
	\label{fig:svDef}
\end{figure}

Prior to delving into the development of the SVO, we present a brief review of the time-derivative of the voltage vector. The voltage vector $\mathbf{v}=V_p(t) e^{j\theta(t)}=V_p(t) e^{j\omega(t) t}$, shown in Fig.~\ref{fig:svDef}, is defined by the instantaneous vector magnitude $V_p(t)$ and the instantaneous angle $\theta (t)$. The rate of change of $V_p(t)$ and the instantaneous speed $\omega (t)$ can be derived from the time derivative of the space vector as 

\begin{equation}\label{eq:derSV}
  \begin{split}
    \frac{d}{dt}\left(\mathbf{v}\right)= \left[j\omega+ \frac{1}{V_p}\frac{d}{dt}(V_p)\right]\mathbf{v}=\Omega \mathbf{v}.
  \end{split}
\end{equation}

\noindent
Here, $(t)$ is dropped from all instantaneous variables in the interest of space and this omission will be followed henceforth. It is worth noting that the imaginary part of $\Omega$ denotes the instantaneous speed of rotation/instantaneous frequency $\omega$ and the real part denotes the normalized rate of change of vector magnitude. We consider a synchronous reference frame aligned with $\mathbf{v}$. The real part $\operatorname{Re\{\Omega \mathbf{u}\}}$ and imaginary part $\operatorname{Im\{\Omega \mathbf{u}\}}$ of the time-derivative are exerted on the original vector $\mathbf{v}$ along the $d$ and $q$ axes, respectively. 

\subsection{Grid Following (GFL) Operation}\label{sec:uocGFL}
We consider a control law on the converter output voltage vector given as   

\begin{equation}\label{eq:uocGFLeq}
  \begin{split}
    \frac{d}{dt}\left(\mathbf{v}\right)= \underbrace{j\omega_0 \mathbf{v}}_{\text{Harmonic oscillator}}\quad + \underbrace{\eta (\mathbf{i_0}-\mathbf{i})e^{j\phi}}_\text{Synchronization feedback},
  \end{split}
\end{equation}

\noindent
where, $\omega_0$ denotes the nominal frequency and  $\eta>0$ is a design parameter; $\phi$ is another design parameter to rotate the current error $\mathbf{e_i}=(\mathbf{i_0}-\mathbf{i})$ for a desired droop relation. The current reference $\mathbf{i_0}=i_{\alpha 0} + j i_{\beta 0}\leftrightarrow i_0=[i_{\alpha0}\ i_{\beta0}]^T$ is obtained using instantaneous power theory\cite{AkagiIPT} as

\begin{equation}\label{eq:iref}
  \begin{split}
 \left[{\begin{array}{c}
         i_{\alpha0}\\
         i_{\beta0}\\
         \end{array}}\right] =\frac{2}{N}\times \frac{1}{V^2_{p}}\left[{\begin{array}{cc}
         v_{\alpha} & v_{\beta}\\
         v_{\beta} & -v_{\alpha}\\
         \end{array}}\right]\left[{\begin{array}{c}
         P_0\\
         Q_0\\
         \end{array}}\right],
  \end{split}
\end{equation}

\noindent
where, $V^2_p=v^2_\alpha+v^2_\beta$; $P_0$ and $Q_0$ denote the real and reactive power set-points/references, respectively and $N$ denotes the number of phases, i.e., $N=1$ and $N=3$ for single and three phase systems, respectively.

Next, we illustrate the principle of operation of \eqref{eq:uocGFLeq}. The first term $j\omega_0\mathbf{v}$ corresponding to the harmonic oscillator, is applied along the $q$-axis and without the synchronization feedback term, \eqref{eq:uocGFLeq} reduces to a simple harmonic oscillator of dimension 2 which results in a voltage vector rotating at $\omega_0$~rad/s with arbitrary magnitude. The synchronization feedback term realigns the voltage vector and adjusts the vector magnitude to track the real and reactive power references. The instantaneous real and reactive power outputs are given as 

\begin{equation}\label{eq:insPQ}
  \begin{split}
 \left[{\begin{array}{c}
         P\\
         Q\\
         \end{array}}\right] =\frac{N}{2}\left[{\begin{array}{cc}
         v_{\alpha} & v_{\beta}\\
         v_{\beta} & -v_{\alpha}\\
         \end{array}}\right]\left[{\begin{array}{c}
         i_\alpha\\
         i_\beta\\
         \end{array}}\right].
  \end{split}
\end{equation}

\noindent
Using \eqref{eq:iref} and \eqref{eq:insPQ}, the current error can be obtained as

{\small
\begin{equation}\label{eq:currErr}
  \begin{split}
    \mathbf{e_i} = \left(\frac{2}{NV^2_p}e_P\right)\mathbf{v}+ j\left(\frac{-2}{NV^2_p}e_Q\right)\mathbf{v}=(e_{iP}+je_{iQ})\mathbf{v},
  \end{split}
\end{equation}
}
\noindent
where, $e_P=P_0-P$ and $e_Q=Q_0-Q$. Evidently, $\mathbf{e_i}$ can be decomposed into two components- a $d$-component $e_{iP}\mathbf{v}$ which corresponds to the error in real power tracking and a $q$-component $je_{iQ}\mathbf{v}$ corresponding to the error in reactive power output. First, we consider $\phi=\pi/2$ and substituting $\mathbf{e_i}$ from \eqref{eq:currErr} into \eqref{eq:uocGFLeq} gives

\begin{equation}\label{eq:uocGFLpw1}
  \begin{split}
    \frac{d}{dt}\left(\mathbf{v}\right)= \left[j(\omega_0+\eta e_{iP})-\eta e_{iQ}\right]\mathbf{v}.
  \end{split}
\end{equation}

\noindent
Comparing \eqref{eq:derSV} and \eqref{eq:uocGFLpw1}, the dynamics of the instantaneous frequency and voltage vector magnitude can be obtained as

\begin{equation}\label{eq:uocGFLpw2}
  \begin{split}
    \frac{d}{dt}(V_p) = -\eta V_p e_{iQ},\qquad &\text{{\small along $d$-axis}};\\
    \omega = \omega_0+\eta e_{iP},\qquad &\text{{\small along $q$-axis}}.
  \end{split}
\end{equation}

\noindent
The oscillator voltage vector magnitude is adjusted until equilibrium is reached along the $d$-axis as $e_{iQ}=0 \iff Q_0=Q$. However, frequency is a network-wide parameter and consequently, along the $q$-axis, a droop response is observed in the real power output as 

\begin{equation}\label{eq:uocGFLpw3}
  \begin{split}
    \omega = \omega_0+\frac{2\eta}{NV^2_P}(P_0-P).
  \end{split}
\end{equation}

\begin{figure}[htb]
	\makebox[\linewidth][c]{\includegraphics[angle = 0, clip, trim=0cm 0cm 0cm 0cm,  width=0.5\textwidth]{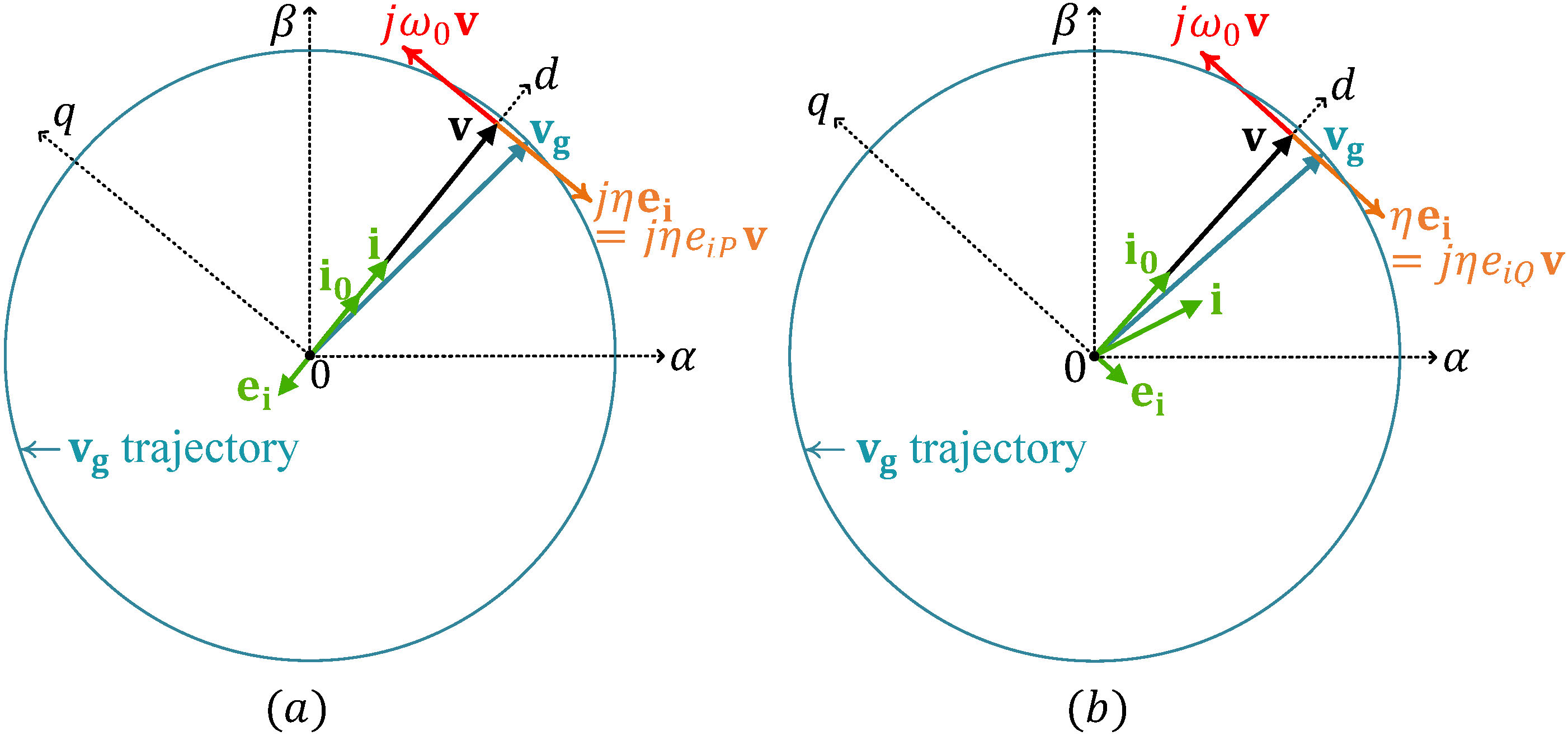}}
	\caption{uVOC GFL operation using \eqref{eq:uocGFLeq} for $P_0\neq 0,\ Q_0=0,\ \omega_g \neq \omega_0$ and - (a) $\phi=\pi/2$, (b) $\phi=0$.}
	\label{fig:svdGFL}
\end{figure}



\noindent
Fig.~\ref{fig:svdGFL}(a) shows an example of converter operation using \eqref{eq:uocGFLeq} for $\phi=\pi/2$, $P_0\neq0$, $Q_0 = 0$, and network/grid frequency $\omega=\omega_g \neq \omega_0$. Using similar analysis, it can be shown that for $\phi=0$ (see Fig.~\ref{fig:svdGFL}(b)), accurate tracking of real power reference is achieved ($e_{iP}=0 \iff P=P_0$) irrespective of network/grid condition whereas a droop response is obtained between reactive power and the instantaneous frequency as

\begin{equation}\label{eq:uocGFLqw3}
  \begin{split}
    \omega = \omega_0-\frac{2\eta}{NV^2_P}(Q_0-Q).
  \end{split}
\end{equation}

The converter cannot maintain the desired output voltage in absence of a grid while using \eqref{eq:uocGFLeq}, instead the converter follows the grid voltage magnitude; hence, this mode of operation is denoted as grid-following (GFL) operation. Accurate tracking of either the real power reference or the reactive power reference can be achieved depending on the choice of $\phi$; integral compensation can be used to achieve accurate tracking of the other. For instance, the real power reference $P_0$ can be generated dynamically by a closed-loop compensator for DC bus voltage regulation in an active-front-end rectifier with $\phi=\pi/2$. DC bus voltage regulation is further discussed in Section~\ref{sec:dcBusReg}. Note that the control law given by \eqref{eq:uocGFLeq} appears similar to stationary frame integral current control \cite{SFcc} utilizing complex signal rotation \cite{vecRot}. However, the distinctions from conventional PI current regulators become evident once the current reference $i_0$ is substituted using \eqref{eq:iref}. The proposed uVOC enables grid synchronization, whereas a PLL is required to generate the current reference for conventional PI regulators. uVOC enables to avoid the synchronization issues of PLL in weak grid conditions. 


\subsection{Grid Forming Operation}\label{sec:uocGFM}
\begin{figure}[htb]
	\makebox[\linewidth][c]{\includegraphics[angle = 0, clip, trim=0cm 0cm 0cm 0cm,  width=0.45\textwidth]{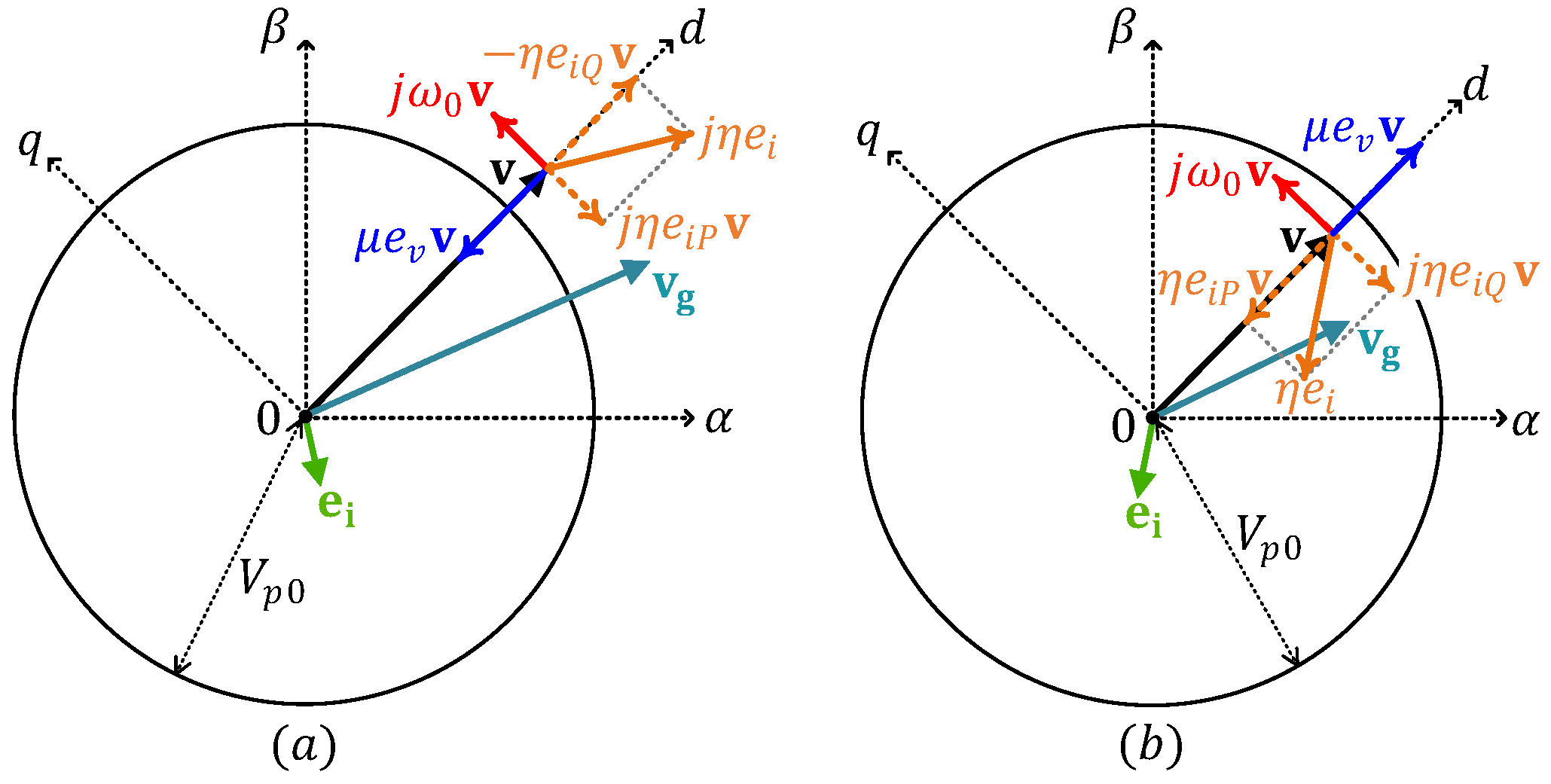}}
	\caption{uVOC GFM operation  - (a) $\phi=\frac{\pi}{2}$, (b) $\phi=0$.}
	\label{fig:svdGFM}
\end{figure}

For GFM operation, the control law on the converter output voltage vector is taken as 

\begin{equation}\label{eq:uocGFMeq}
  \begin{split}
    \frac{d}{dt}\left(\mathbf{v}\right)= \underbrace{j\omega_0 \mathbf{v}}_{\text{Har. osc.}} + \underbrace{\mu(V^2_{p0}-V^2_p)\mathbf{v}}_{\text{Magnitude correction}} + \underbrace{\eta (\mathbf{i_0}-\mathbf{i})e^{j\phi}}_\text{Sync. feedback},
  \end{split}
\end{equation}

\noindent
where, $\mu>0$ is a design parameter and $V_{p0}$ denotes the nominal set-point for the voltage vector magnitude $V_p$. The control law given by \eqref{eq:uocGFMeq} is built on the synchronization results reported in \cite{dVOC1,dVOC2,dVOC3,dVOC4}. The principle of operation can be explained using the simplified system shown in Fig.~\ref{fig:simSysDes}. The magnitude correction term is exerted radially along the $d$-axis on the voltage vector and it is opposed by either the error in real power output or the reactive power output depending on the selection of $\phi$. To illustrate the instantaneous droop response, first we consider $\phi=\pi/2$; {Fig.~\ref{fig:svdGFM}(a)} shows the corresponding operation. Without loss of generality, an arbitrary current error $\mathbf{e_i}$ is considered.  Using \eqref{eq:currErr}, the controller given by \eqref{eq:uocGFMeq}, can be rearranged as

\begin{equation}\label{eq:uocGFMeq1}
  \begin{split}
    \frac{d}{dt}\left(\mathbf{v}\right)= \left[j\left(\omega_0+\eta e_{iP}\right)+\left(\mu e_v-\eta e_{iQ}\right)\right]\mathbf{v}.
  \end{split}
\end{equation}

\noindent
Here, $e_v=V^2_{p0}-V^2_p$. The dynamics along the $d$ and $q$ axes are obtained as 

\begin{equation}\label{eq:uocGFMpw1}
  \begin{split}
    \frac{d}{dt}(V_p) = \mu V_p(V^2_{p0}-V^2_p)-\eta V_p e_{iQ},\qquad \text{{\small along $d$-axis}};\\
    \omega = \omega_0+\eta e_{iP},\qquad \text{{\small along $q$-axis}}.
  \end{split}
\end{equation}

\noindent
Along the $d$-axis, unlike GFL operation, the voltage vector magnitude cannot be adjusted freely due to the magnitude correction term; instead, a nonlinear droop in the instantaneous magnitude of the output voltage vector is observed which can be obtained by setting $\frac{d}{dt}(V_p)=0$ as

\begin{equation}\label{eq:uocDRqv}
  \begin{split}
    V^2_p &= V^2_{p0}+\frac{2\eta}{\mu N V^2_p}(Q_0-Q).
  \end{split}
\end{equation}

\noindent
It is worth noting that the non-linearity in the droop response becomes prominent when the system operates far away from the nominal set-points $Q_0$ and $V_{p0}$; close to the nominal condition, the droop relation can be approximated as linear. The closed-form solution for the steady-state voltage is listed in Appendix~\ref{apndx:a}. Along the $q$-axis, $P-\omega$ droop, same as \eqref{eq:uocGFLpw3}, is obtained. It is worth noting that the droop response is achieved naturally by the time-domain implementation without explicit regulation of voltage amplitude and frequency using the calculated output power which is required in droop based methods\cite{vocVsDc1}.



Using similar analysis, it can be shown that $P-V$ and $Q-\omega$ droop responses are obtained  along the $d$ and $q$ axes, respectively, for $\phi=0$. Fig.~\ref{fig:svdGFM}(b) shows an example of GFM operation with $\phi=0$ for an arbitrary current error $\mathbf{e_i}$ and the infinite bus voltage $\mathbf{v_g}$. Coupled droop relations among $P,Q,\omega,$ and $V$ can be achieved using $\phi\in (0, \pi/2)$. However, for faster synchronization $\phi=0$ and $\phi=\pi/2$ should be used in dominantly resistive and inductive networks, respectively. The detail derivation of instantaneous droop response for GFM operation for arbitrary rotation angle $\phi$ is provided in Appendix~\ref{apndx:a}. In essence, the harmonic oscillator term tends to restore the nominal frequency by forcing the voltage vector $\mathbf{v}$ along the $q$-axis, whereas, the magnitude correction term forces the voltage vector along the $d$-axis to restore the nominal magnitude; however, both are opposed by the synchronization feedback term based on the instantaneous real and reactive power output of the converter. Consequently, a self synchronizing system is obtained.

\begin{figure}[htb]
	\makebox[\linewidth][c]{\includegraphics[angle = 0, clip, trim=0cm 0cm 0cm 0cm, width=0.5\textwidth]{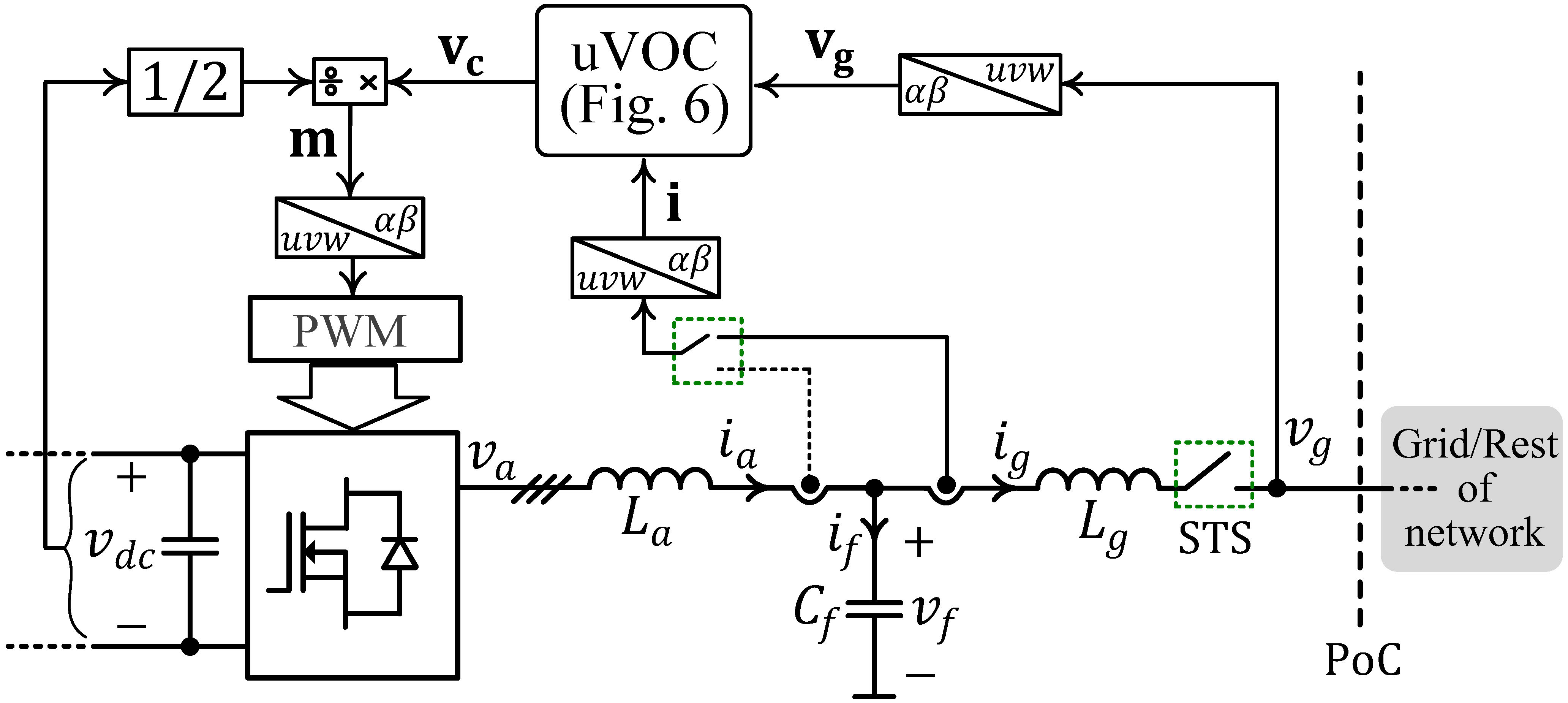}}
	\caption{A three-phase VSC based on uVOC using grid-side current feedback; alternatively, converter-side current feedback can be used for uVOC implementation.}
	\label{fig:sysDes}
\end{figure}

\section{Implementation of uVOC}\label{sec:ImpIssues}
\begin{figure}[htb]
	\makebox[\linewidth][c]{\includegraphics[angle = 0, clip, trim=0cm 0cm 0cm 0cm, width=0.5\textwidth]{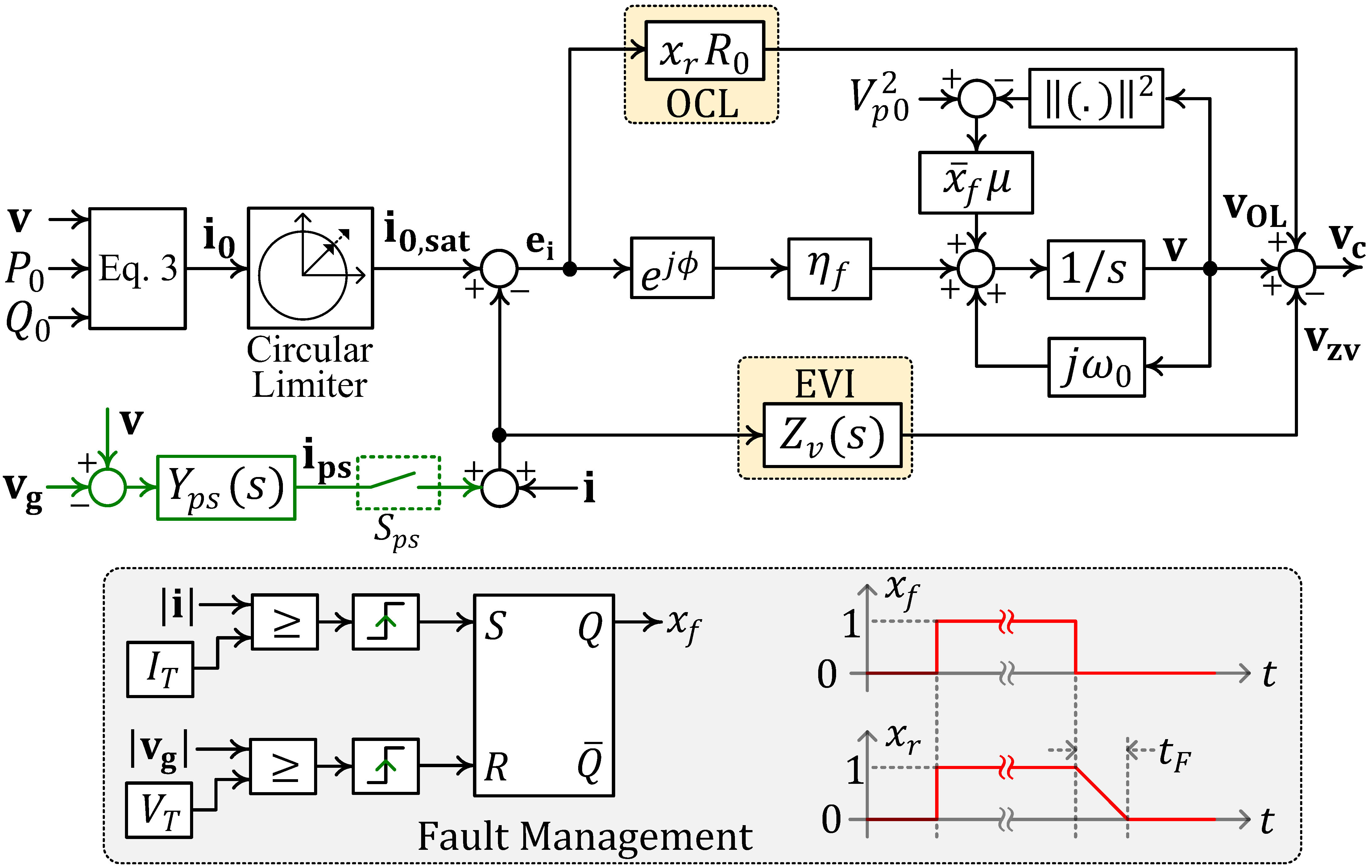}}
	\caption{uVOC implementation; controller components corresponding to pre-synchronization are marked in green.}
	\label{fig:uVOC}
\end{figure}

A three-phase voltage source converter (VSC) using uVOC is shown in Fig.~\ref{fig:sysDes}. The switching duty-ratio $\mathbf{m}$ for the PWM is generated from uVOC output $\mathbf{v_c}$ using appropriate scaling based on the DC bus voltage measurement. The controller, shown in Fig.~\ref{fig:uVOC}, is implemented based on space vectors in stationary $\alpha\beta$-reference frame; either of the converter-side current or the network/grid-side current feedback can be used for controller implementation. GFM operation is used in grid-supporting applications or for decentralized power sharing in an islanded microgrid. GFL operation should be used for interfacing non-dispatchable sources or loads, such as motor drives. Identical control structure is used for GFL and GFM applications; the controller can transition from GFM to GFL operation by simply assigning $\mu=0$. The real power reference $P_0$ is dynamically generated if DC bus voltage regulation is desired in GFL applications. The pre-synchronization filter, marked in green in Fig.~\ref{fig:uVOC}, is used for soft start-up and transition from islanded to grid-tied operation. The emulated virtual impedance (EVI) is used in all modes of operation. The fault management block generates two signals $x_f$ and $x_r$ to enable fault ride-through operation. During normal operation, defined by $\{x_f=0, x_r=0\}$, the controller shown in Fig.~\ref{fig:uVOC}, excluding the EVI and the pre-synchronization components, effectively implements \eqref{eq:uocGFLeq} and \eqref{eq:uocGFMeq} for GFL and GFM operations, respectively. Note that the complementary signal is defined as $\overline{x}_f=\operatorname{NOT}\{x_f\}$. The circular limiter generates saturated current reference as

\begin{equation}\label{eq:cirLim}
  \begin{split}
    \mathbf{i_{0,sat}}=
    \begin{cases}
        \mathbf{i_0}, & |\mathbf{i_0}|\leq I_{m}\\
        \mathbf{i_0}\times (I_m/|\mathbf{i_0}|), & \text{otherwise}.
    \end{cases}
  \end{split}
\end{equation}
\noindent
Here, $I_m$ denotes the maximum allowable current. Under normal operating conditions the circular limiter block is transparent and during fault, marked by $|\mathbf{i_0}|>I_m$, the current reference vector is radially limited, i.e., magnitude is limited keeping the angle unaltered, which enables synchronization during fault. In the following subsections, first, SVO parameter selection guidelines are developed; the EVI, single-phase operation, and the fault management block are explained subsequently. Detailed implementation of the pre-synchronization filter is presented in Section~\ref{sec:preSync} after we develop the small-signal model of uVOC. 

\subsection{SVO Parameter Selection}\label{sec:paramSelec}
\begin{figure}[htb]
	\makebox[\linewidth][c]{\includegraphics[angle = 0, clip, trim=0cm 0cm 0cm 0cm,  width=0.5\textwidth]{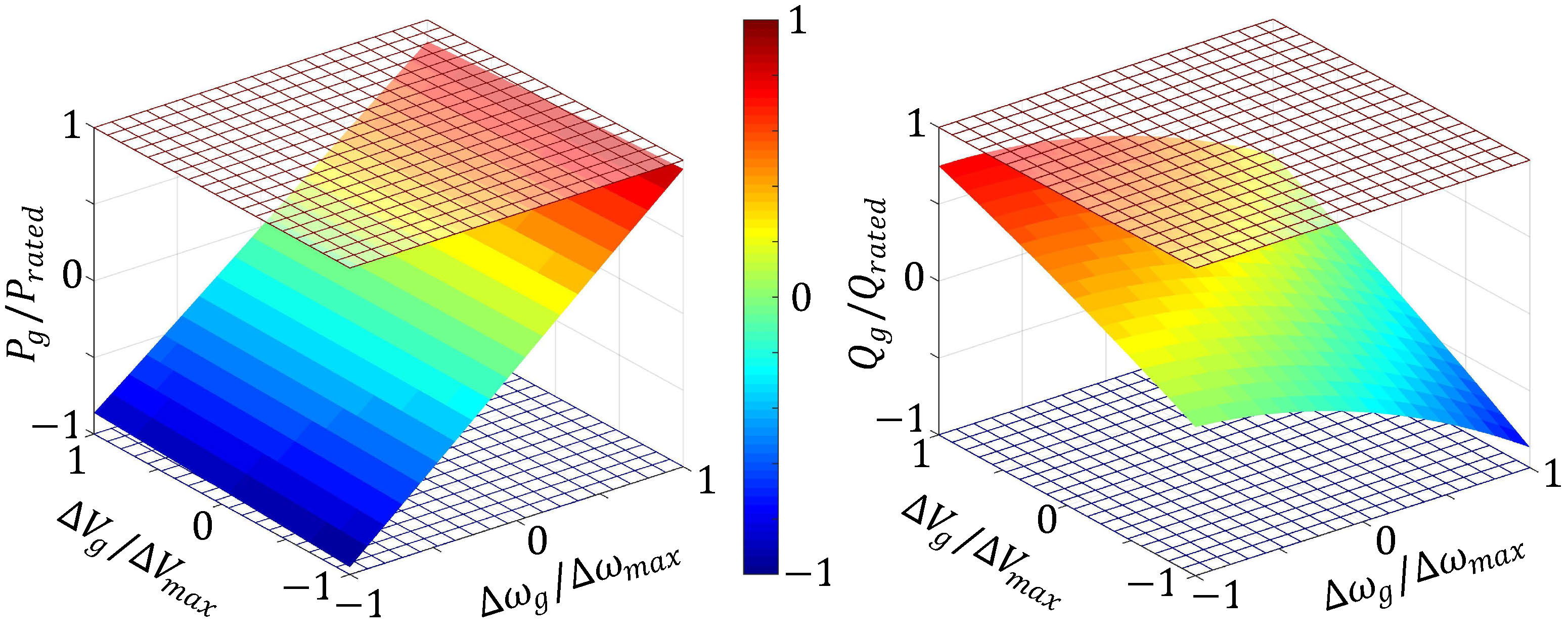}}
	\caption{Real and reactive power outputs of the VSC at PoC are limited within rated values for frequency and voltage variation at the PoC over the entire operating range.}
	\label{fig:PQWV}
\end{figure}

The SVO parameters $\eta$ and $\mu$ are selected based on the desired steady-state response. For GFM operation, the parameter selection process begins with the specification such that the output power of the VSC should be limited to $[-P_{rated},\ P_{rated}]$ and $[-Q_{rated},\ Q_{rated}]$ when the voltage and frequency deviation at the PoC varies within the full range, i.e., $\Delta V_g=(V_0-V_g)\in [-\Delta V_{max}, \Delta V_{max}]$ and $\Delta \omega_g=(\omega_0-\omega_g)\in [-\Delta \omega_{max}, \Delta \omega_{max}]$, respectively, where $V_0=V_{p0}/\sqrt(2)$ and $V_g=V_{gp}/\sqrt(2)$ denote the respective L-N RMS values. Detail derivation of the design method is provided in Appendix~\ref{apndx:b}. Using~\eqref{eq:wmax} to~\eqref{eq:etaMu1}, the selection of parameter values are summarised in Table~\ref{TB:paramSelec}. The analysis is based on the steady-state droop relation at the poles of the switch network. Due to the LCL filter, the droop response at the PoC differs from that at the poles of switch network. However, the design guidelines listed in Table~\ref{TB:paramSelec} ensure that the real and reactive power output at the VSC terminal at PoC are limited within the rated values for variation of network frequency and voltage over the full allowable range. To illustrate the parameter selection process, we consider the three-phase VSC listed in Table~\ref{TB:vscParam}. The passive filter is designed to achieve $>65$dB ripple attenuation at the switching frequency which restricts the ripple current below $0.2\%$ of the rated current. For $\Delta V_{max}=0.05V_0$, $\Delta \omega_{max}=\pi\ \text{rad/s}$, and $\phi=\pi/2$, $\eta=16.6253$ and $\mu=5.2029\times 10^{-4}$ are obtained. For network frequency and PoC voltage variation in the full range, the VSC output real and reactive powers are shown in Fig.~\ref{fig:PQWV}. Evidently, over the entire operating range, the converter output powers are limited within the rated values. The results in Fig.~\ref{fig:PQWV} are obtained numerically from power flow solution for the LCL filter and the steady-state response of SVO given by \eqref{eq:qAxisDyn} and \eqref{eq:opV} in Appendix~\ref{apndx:a}. Selection of $\eta$ for GFL operation is done following identical steps as GFM operation. 

\renewcommand{\arraystretch}{1}
\begin{table}[htb]
\centering
\caption{SVO Parameter Design}
\begin{tabular}{ccc}
\toprule
\makecell[ct] Parameter & For $\phi=\pi/2$ & For $\phi=0$ \\
\midrule
\addlinespace[5pt]
$\displaystyle \eta$ & $\displaystyle \frac{N\Delta\omega_{max}V^2_{max}}{P_{rated}}$ & $\displaystyle \frac{N\Delta\omega_{max}V^2_{max}}{Q_{rated}}$\\
\addlinespace[5pt]
$\displaystyle \mu$ & $\displaystyle \frac{2\eta Q_{rated}}{N[(2V^2_{max}-V^2_0)^2-V^4_0]}$ & $\displaystyle \frac{2\eta P_{rated}}{N[(2V^2_{max}-V^2_0)^2-V^4_0]}$\\
\addlinespace[5pt]
\bottomrule
\end{tabular}
\label{TB:paramSelec}
\vspace{-15pt}
\end{table}
\renewcommand{\arraystretch}{1}

\renewcommand{\arraystretch}{1}
\begin{table}[htb]
\centering
\caption{Voltage Source Converter Ratings}
\begin{tabular}{ p{1cm}p{4cm}p{1.5cm}}
\hline
\hline
$S_{rated}$ & Rated power & $10\ \text{kVA}$\\
$P_{rated}$ & Rated real power & $9\ \text{kW}$\\
$Q_{rated}$ & Rated reactive power & $4.4\ \text{kVAR}$\\
$V_0$ & Nominal (L-N RMS) voltage & $120\ \text{V}$\\
$\omega_0$ & Nominal frequency & $2\pi(60)\ \text{rad/s}$\\
$f_{sw}$ & Switching frequency & $10$kHz\\
$f_s$ & Sampling frequency & $10$kHz\\
$L_a$ & Converter-side inductor & $7.78\ \%\text{pu}$\\
$L_g$ & Network-side inductor & $5.24\ \%\text{pu}$\\
$C_f$ & Filter capacitor & $8.79\ \%\text{pu}$\\
\hline
\hline
\end{tabular}
\label{TB:vscParam}
\vspace{-15pt}
\end{table}

\subsection{Emulated Virtual Impedance (EVI)}\label{sec:EVI}
EVI is used in all modes of operation and has two key purposes, namely, harmonic compensation and stabilization of SVO. The converter output impedance can be selectively increased to very high values at the harmonic frequencies to suppress harmonic distortion in converter output current originating from non-ideal effects such as dead-time/blanking time of power-devices and/or harmonic distortion in network/grid voltage. EVI is achieved as 

\begin{equation}\label{eq:hcs}
  \begin{split}
 v_{zvx}= Z_v(s)&\times i_{x},\quad \forall x\in \{\alpha,\ \beta\};\\
 Z_v(s) = \frac{R_{vir}}{(s/\omega_c+1)} +& \sum_h \frac{-K_h \omega_{B,h} \omega_h}{s^2 + \omega_{B,h}s + \omega^2_h},\quad \text{for } \mathbf{i}=\mathbf{i_g};\\
 Z_v(s) = \frac{R_{vir}}{(s/\omega_c+1)} +& \sum_h \frac{K_h \omega_{B,h} s}{s^2 + \omega_{B,h}s + \omega^2_h};\quad \text{for } \mathbf{i}=\mathbf{i_a},
  \end{split}
\end{equation}

\noindent
where, $K_h$ and $\omega_{B,h}$ denote the desired impedance magnitude and the bandwidth of the resonant filter at harmonic frequency $\omega_h$, respectively. Note that the resonant parts of $Z_v(s)$ are designed to be inductive or resistive for harmonic suppression when grid-side current or converter-side current feedback is used, respectively. Detail design guidelines for harmonic compensation can be found in \cite{TpelHCS}. Virtual impedance design guidelines to compensate harmonic voltage distortion at the PoC while feeding nonlinear loads can be found in \cite{hvsApec}. However, to ensure dynamic stability of the SVO, a virtual resistance $R_{vir}$ with a limited bandwidth of $\omega_c$ is used in both cases. The control laws given by \eqref{eq:uocGFLeq} and \eqref{eq:uocGFMeq} are of integral form; the lack of proportional term leads to poor damping. In essence, the virtual resistance $R_{vir}$ constitutes a proportional compensation and provides the necessary damping. The damping effect of $R_{vir}$ through small-signal analysis is illustrated in Section~\ref{sec:rSelec}.


\subsection{Single Phase Operation}
uVOC is a vector controller running on both $\alpha$ and $\beta$ axis. In a single-phase implementation, the full vector controller is used. However, the virtual impedance $Z_{v}(s)$ emulation is required only for the $\alpha$ axis. The feedback signal $i_\beta$ can be generated by delaying the actual converter output current $i_\alpha=i_g$ or $i_a$ by $T_0/4=2\pi/(4\omega_0)$ and the modulating signal for PWM is obtained from $v_{c\alpha}$. All analysis and design guidelines presented in the following sections are generalized to apply identically for single phase and three phase applications using the parameter $N$. 

\subsection{Fault Management}\label{sec:FMT}
A \emph{fault state} is latched, i.e., $x_f$ is pulled \emph{high}, once $|\mathbf{i}|>I_T$ is detected, where $I_T$ denotes the over-current threshold. An over-current may be triggered in a GFM converter by an overload or an AC fault while serving local loads; a GFM or GFL converter can be subjected to over-current in grid-tied operation due to an upstream fault or voltage sag at the PoC. The \emph{fault state} is cleared, i.e., $x_f$ is pulled \emph{low}, when the terminal voltage $v_g$ returns above the low-voltage threshold $V_T$. Once a fault is detected in a GFM converter, the magnitude correction term is disabled by the complementary signal $\overline{x}_f$. For over-current limiting (OCL), a series compensation is applied as 

\begin{equation}\label{eq:VOCL}
  \begin{split}
    \mathbf{v_{OL}}=x_r R_{0}(\mathbf{i_{0,sat}}-\mathbf{i}).
  \end{split}
\end{equation}

\noindent
The OCL gain is selected as $R_{0}=\omega_{OCL}(L_a+L_g)$, where $\omega_{OCL}$ denotes the desired current control bandwidth. The OCL compensation is critical in limiting current transients. In addition to the duration of the fault, large current transients are likely immediately after the fault is cleared. To facilitate smooth transition into normal operation once the fault is cleared, the OCL compensation is disabled gradually using a ramp in $x_r$ over a short duration $t_f$. The small signal model presented in Section~\ref{sec:ssmFRT} reveals that the OCL compensation moves the dominant poles of the system very close to the imaginary axis leading to slower synchronization response. To circumvent such sluggish response, the synchronization gain is augmented as $\eta_f=\eta\{1+x_f (R_0/\tau_f)\}$ during fault ride-through; the selection of $\tau_f$ can be made through small signal analysis to achieve a desired settling time. In fault mode, the reactive power reference $Q_0$ can be increased to provide voltage support at the PoC while the converter output current remains clamped at the maximum allowable value.


\section{Small Signal Model of uVOC Based VSC}\label{sec:ssm}
Small-signal modeling and analysis is necessary for the selection different control parameters such as $R_{vir}$ and the design of DC bus voltage regulator. 

\subsection{Normal Operation}
\begin{figure*}[b]
	\makebox[\linewidth][c]{\includegraphics[angle = 0, clip, trim=0cm 0cm 0cm 0cm,  width=1\textwidth]{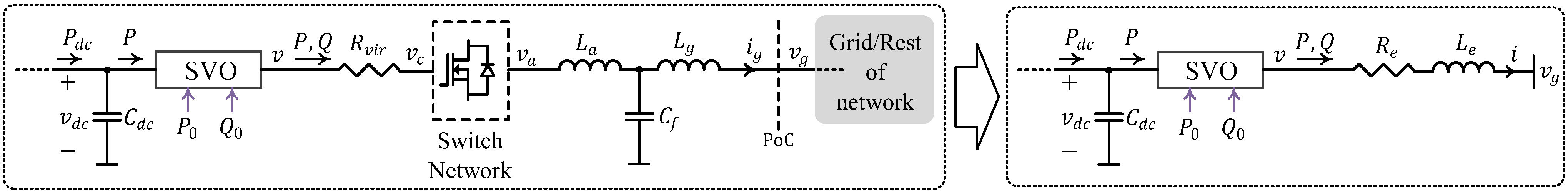}}
	\caption{Equivalent model of a uVOC based VSC for small signal analysis.}
	\label{fig:equiMod}
\end{figure*}

During normal operation, the circular limiter is transparent. An equivalent model, shown in Fig.~\ref{fig:equiMod}, of the VSC can be constructed for small signal analysis. The SVO controls the power flow between the DC bus and the AC terminal. The virtual resistance $R_{vir}$ effectively appears between the oscillator and the switch network which is necessary to ensure stability of the SVO regardless of the electromagnetic line-dynamics owing to variation in the grid-impedance. In the frequency range of interest, the filter capacitor, the controller implementation delay, and the effect of the PWM can be ignored. The total equivalent series resistance including $R_{vir}$, any parasitic resistance in the LCL filter, and the resistive element in the grid impedance $Z_N=R_N+sL_N$ are combined as $R_e$ and the total equivalent inductance is taken as $L_e=L_a+L_g+L_N$. In a synchronous reference frame rotating at $\omega_*$ rad/s, the SVO dynamics given by \eqref{eq:dAxisDyn} and \eqref{eq:qAxisDyn} in Appendix~\ref{apndx:a}, can be rewritten as-

\small
\begin{equation}\label{eq:polOscSRF}
  \begin{split}
  \Dot{V} &= 2\mu V (V^2_0-V^2) + \frac{\eta}{NV}\left[(P_0-P)\cos{\phi}+(Q_0-Q)\sin{\phi}\right];\\
  \Dot{\theta}_s &= \omega_0 - \omega_* + \frac{\eta}{NV^2}\left[(P_0-P)\sin{\phi} - (Q_0-Q)\cos{\phi}\right].
  \end{split}
\end{equation}
\normalsize
\noindent
Here, $\theta(t)=\int \omega dt=\omega_* t+\theta_s(t)$. Without loss of generality, the network frequency can be taken as $\omega_* $ and the rotating reference frame can be taken to be aligned with $\mathbf{v_g}$. Note that the synchronous reference frame used in Section~\ref{sec:svo} is aligned with the oscillator voltage vector $\mathbf{v}$. In the synchronous $dq$-frame, the dynamics of the converter output current $\mathbf{i_{dq}}$  can be derived as

\small
\begin{equation}\label{eq:iDyn}
  \begin{split}
 \left[{\begin{array}{c}
        \Dot{I}_{d}\\
         \Dot{I}_{q}\\
         \end{array}}\right] =\left[{\begin{array}{cc}
         -\frac{R_e}{L_e} & \omega_*\\
         -\omega_* & -\frac{R_e}{L_e}\\
         \end{array}}\right]\left[{\begin{array}{c}
         I_{d}\\
         I_{q}\\
         \end{array}}\right] + \frac{1}{L_e}\left[{\begin{array}{c}
         V_a\cos{(\theta_s)}-V_{g}\\
         V_a\sin{(\theta_s)}\\
         \end{array}}\right].
  \end{split}
\end{equation}
\normalsize

\noindent
Here, $\mathbf{i_{dq}}=i_d+j i_q=(I_d+jI_q)/\sqrt{2}$. For the implementation shown in Fig.~\ref{fig:sysDes}, $V_a=V$ can be assumed since the instantaneous feedback of the DC bus voltage is used to determine the modulating duty ratio $\mathbf{m}$. However, for real applications, measurement noise propagates directly into the modulating signal if instantaneous DC bus voltage feedback is used; instead, the nominal DC bus voltage reference $V^*_{dc}$ can be used which gives $V_a = (v_{dc}/V^*_{dc})V$. The real power $P$ and reactive power $Q$ are given as 

\begin{equation}\label{eq:PQSRF}
  \begin{split}
 \left[{\begin{array}{c}
         P\\
         Q\\
         \end{array}}\right] =\frac{N V v_{dc}}{V^*_{dc}}\left[{\begin{array}{cc}
         \cos{(\theta_s)} & \sin{(\theta_s)}\\
         \sin{(\theta_s)} & -\cos{(\theta_s)}\\
         \end{array}}\right]\left[{\begin{array}{c}
         I_{d}\\
         I_{q}\\
         \end{array}}\right].
  \end{split}
\end{equation}

\noindent
The DC bus voltage dynamics can be derived as

\begin{equation}\label{eq:dcBusDyn}
  \begin{split}
  \frac{d}{dt}\left(\frac{1}{2}C_{dc}v^2_{dc}\right) = P_{dc} - P.
  \end{split}
\end{equation}
Using \eqref{eq:polOscSRF}, \eqref{eq:iDyn}, \eqref{eq:PQSRF}, and \eqref{eq:dcBusDyn}, the small signal model can be obtained as 

\renewcommand{\arraystretch}{0.9}
\begin{equation}\label{eq:ssMod}
  \begin{split}
  \arraycolsep=1.4pt\def\arraystretch{1}
 \left[{\begin{array}{c}
         \Delta \Dot{x}\\
         \Delta \Dot{v}_{dc}\\
         \end{array}}\right] 
         =\left[{\begin{array}{cc}
         A_{11} & A_{12}\\
         A_{21} & A_{22}\\
         \end{array}}\right]\left[{\begin{array}{c}
         \Delta x\\
         \Delta v_{dc}\\
         \end{array}}\right] + \left[{\begin{array}{c}
         B_{11}\\
         B_{21}\\
         \end{array}}\right]\left[{\begin{array}{c}
         \Delta P_0\\
         \Delta Q_0\\
         \end{array}}\right],
  \end{split}
\end{equation}

\noindent
where, $x=[\Delta I_d\ \Delta I_q\ \Delta V\ \Delta\theta_s]^T$ and the detail forms of $A_{11}\in \mathbb{R}^{4 \times 4}$, $A_{21}\in \mathbb{R}^{1 \times 4}$, $A_{12}\in \mathbb{R}^{4\times 1}$, $A_{22}\in \mathbb{R}$, $B_{11}\in \mathbb{R}^{4 \times 2}$, and $B_{21}\in \mathbb{R}^{1 \times 2}$ are provided in Appendix~\ref{apndx:d}. Note that $\Delta$ denotes small perturbation around the operating point.  

\subsection{Fault Ride-Through Operation}\label{sec:ssmFRT}
During a fault, the circular limiter saturates the current reference and current error as 

\begin{equation}\label{eq:iSatFF1}
  \begin{split}
    \mathbf{i_{0,sat}} = (I_m/|\mathbf{i_0}|) \mathbf{i_0}=&(K_m V) \mathbf{i_0};\\\mathbf{e_i}=1/(N V^2)\{(K_m V P_0-P)-&j(K_m V Q_0-Q)\}\mathbf{v}, 
  \end{split}
\end{equation}

\noindent
where $K_m = N I_m/(\sqrt{2(P_0^2+Q_0^2)})$. Now the modified SVO dynamics can be obtained by simply substituting $P_0$ and $Q_0$ in \eqref{eq:polOscSRF} by $K_m V P_0$ and $K_m V Q_0$, respectively. Due to the over-current limiting compensation (see Fig.~\ref{fig:uVOC}), the effective resistance increases as $R_e=R_{vir}+R_0$ and an additional term $+R_0 \mathbf{i_{0,sat}}$ is added on the right side of \eqref{eq:iDyn}. The output current dynamics in the synchronous frame can be derived as 

\begin{equation}\label{eq:iSatFF2}
  \begin{split}
    \mathbf{i_{0,sat}}=(K_m/N)(P_0-jQ_0)(\cos{\theta_s} +j\sin{\theta_s}). 
  \end{split}
\end{equation}

\noindent
Following similar steps as the preceding subsection, the small signal model of the system during fault mode operation can be derived.

\section{Control Design for Dynamic Performance}\label{sec:dynParamSelec}
The virtual resistance $R_{vir}$, DC bus voltage regulator, and the pre-synchronization filter are designed to ensure the desired dynamic response.  

\subsection{Selection of Virtual Resistance}\label{sec:rSelec}
The small-signal model is used to select $R_{vir}$ once the SVO parameters are chosen. For GFM operation, we assume that the DC bus voltage is maintained by a DC source or regulated by a AC/DC or DC/DC converter stage; therefore, the DC bus dynamics is ignored. The linearized system dynamics for the GFM operation is given by

\begin{equation}\label{eq:ssGFM}
  \begin{split}
    \Dot{x}=A_{11}x + B_{11}
    \left[{\begin{array}{cc}
         \Delta P_0 & \Delta Q_0
         \end{array}}\right]^T.
  \end{split}
\end{equation}

\noindent
The system poles are given by $\lambda(A_{11})$, where $\lambda(.)$ denotes eigenvalue. To illustrate the selection of $R_{vir}$, we consider the VSC listed in Table~\ref{TB:vscParam} to be connected to a stiff grid with a dominantly inductive network/grid impedance $Z_N\approx sL_N, L_N=1\text{mH}\equiv 8.7\%\ \text{pu}$. To account for the worst-case from stability perspective, the LCL filter elements are assumed to be lossless. Table~\ref{TB:varRe} lists the system poles for different values of the equivalent series resistance $R_e\approx R_{vir}$. For $R_{vir}=0.5\%$, the system becomes unstable, whereas for $R_{vir}=1.15\%$ a stable, yet very lightly damped system results which is evident from the real part of the complex-conjugate pole pair. Lastly, a well-damped system response is achieved for $R_{vir}=4.9\%$. Following similar steps, $R_{vir}$ can be selected for GFL operation using \eqref{eq:ssGFM} for $\mu=0$.

\renewcommand{\arraystretch}{1}
\begin{table}[h]
\centering
\caption{System Poles for Different Values of $R_{vir}$}
\begin{tabular}{c|ccc}
\toprule
$R_e\approx R_{vir}$ & \multicolumn{3}{c}{$\lambda(A_{11})$}\\
\midrule
$0.5\%\ \text{pu}$ & $9.16\pm378.12j$ & $-47.57$ & $-17.90$\\
$1.15\%\ \text{pu}$ & $-1.94\pm377.6j$ & $-47.72$ & $-17.91$\\
$4.9\%\ \text{pu}$ & $-66.61\pm374.56j$ & $-47.61$ & $-17.68$\\
\bottomrule
\end{tabular}
\label{TB:varRe}
\vspace{-15pt}
\end{table}

\subsection{DC Bus Voltage Regulation (GFL Operation)}\label{sec:dcBusReg}
DC bus voltage regulation can be achieved by using the GFL form of uVOC and dynamically generating $P_0$ using a closed-loop voltage regulator. To design the DC bus voltage regulator, the uncompensated open-loop response $G_{OL}(s)=\Delta v_{dc}(s)/(-\Delta P_0(s))$ can be derived from the transfer function as

\begin{equation}\label{eq:ssRecMode}
  \begin{split}
    G_{rec}(s)= C_{rec}\left(s\mathbb{I}-A\right)^{-1}B,
  \end{split}
\end{equation}

\noindent
where, $G_{rec}(s)\in \mathbb{C}^{1\times2}$ and $\mathbb{I}$ denotes the identity matrix of dimension $5$, $C_{rec}=[0\ 0\ 0\ 0\ 1]$. The definitions of $A$ and $B$ are provided in Appendix~\ref{apndx:d}. A lead-lag filter combined with a PI compensator can be used for closed-loop regulation, given as follows:

\begin{equation}\label{eq:Fdc}
  \begin{split}
    F_{dc}(s)= K_{pdc}\left(1+\frac{1}{sT_{i}}\right)\times \sqrt{\frac{\omega_p}{\omega_z}}\left(\frac{s+\omega_z}{s+\omega_p}\right).
  \end{split}
\end{equation}

\noindent
Here, $\omega_p$ and $\omega_z$ for the lead-lag filter should be chosen to achieve sufficient phase boost at the gain cross-over frequency \cite{ErikssonBook}, whereas the integral time-constant $T_i$ should be chosen for sufficient DC gain through bode plot analysis.



\subsection{Pre-Synchronization}\label{sec:preSync}
The pre-synchronization filter (see Fig.~\ref{fig:uVOC}) is essentially a first-order low-pass filter in the form of a virtual RL branch taken as 

\begin{equation}\label{eq:ps}
  \begin{split}
    \mathbf{i_{ps}}=Y_{ps}(s)(\mathbf{v}-\mathbf{v_g});\quad Y_{ps}(s)=1/(sL_{ps}+R_{ps})
  \end{split}
\end{equation}

\noindent
The parameters can be chosen as $L_{ps}\approx (L_a+L_g)$ and $R_{ps}\approx R_{vir}$. It is worth noting that exact knowledge of the LCL filter parameters are not required for the parameter selection. To illustrate the functionality, two distinct use-cases can be defined for the pre-synchronization filter. 

\subsubsection{STS Closing for GFM Converters Serving Local Loads}
For GFM converters while serving local loads, the voltages across the STS need to be synchronized prior to closing the STS. In such scenarios, the pre-synchronization block is run to generate a virtual current $i_{ps}$, which is added to the actual converter output current and the resultant total current is used as feedback to the SVO. The virtual current $i_{ps}$ gives an estimate of the current that would flow between the VSC and the network at the PoC if the STS were closed. Due to the virtual current $i_{ps}$ feedback, SVO adjusts the oscillator voltage vector to minimize the virtual real and reactive power flow. When the amplitude of the virtual current $|i_{ps}|$ stabilizes, the STS can be closed safely.

\subsubsection{Start-Up in GFL Converters}\label{sec:psGFL}
For GFL operation, the STS can be closed before the switching of the power devices are initiated. For instance, during start-up of an active-front-end rectifier, the DC bus may be charged by using the switch network as an uncontrolled diode bridge. Meanwhile, the SVO can be synchronized with the measured voltage $\mathbf{v_g}$ using the pre-synchronization filter and once $|i_{ps}|$ stabilizes, the DC bus voltage control loop and the switching of the power devices can be initiated without large transients.

\section{Simulation Results}\label{sec:simRes}
All simulations are performed in \emph{PLECS standalone} using detailed switching model of a three-phase VSC with parameters listed in Table~\ref{TB:vscParam}.

\subsection{Small Signal Model Validation}\label{sec:anaModVal}
\begin{figure}[htb]
	\makebox[\linewidth][c]{\includegraphics[angle = 0, clip, trim=0cm 0cm 0cm 0cm,  width=0.425\textwidth]{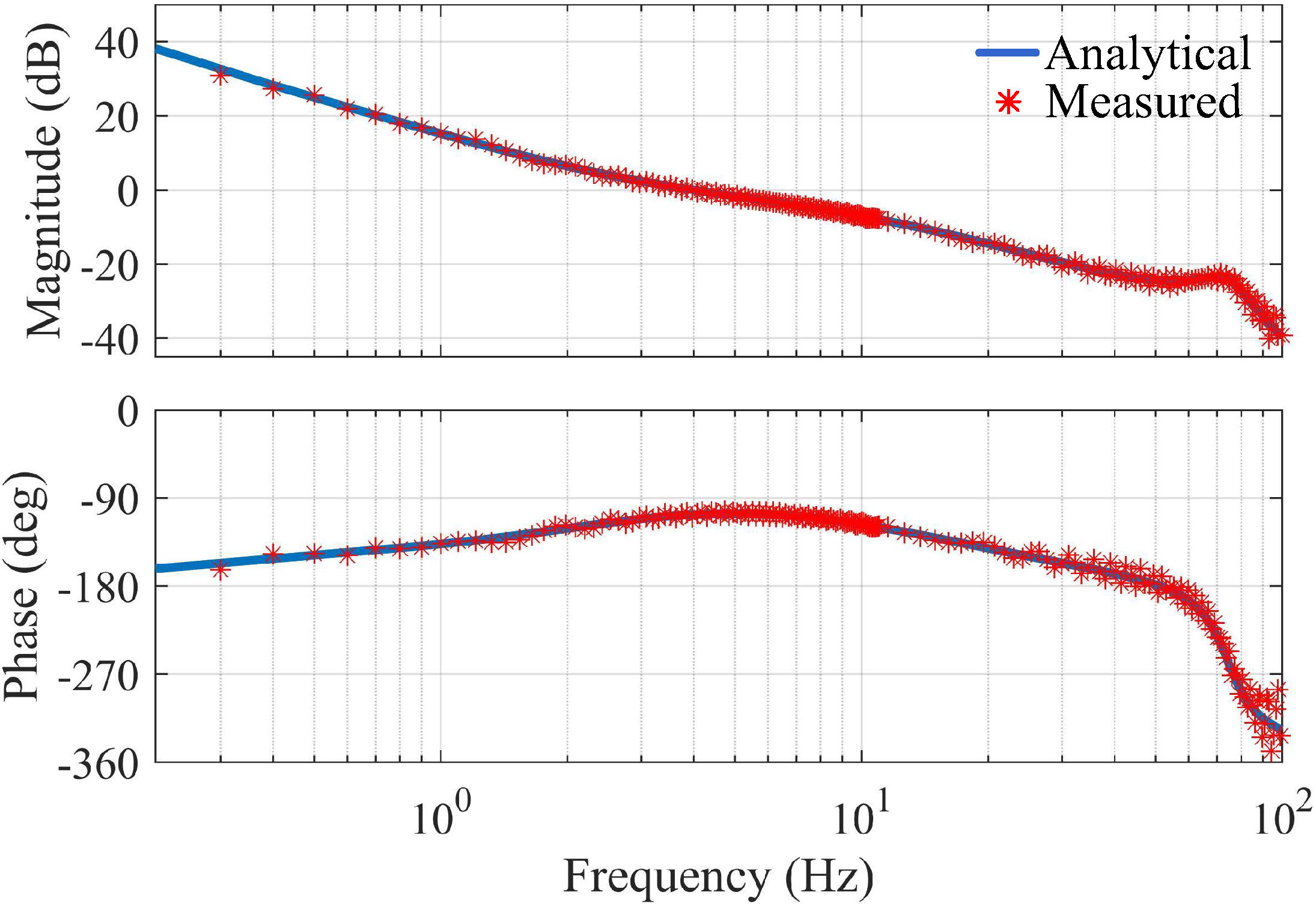}}
	\caption{Comparison of analytical and measured frequency responses ($F_{dc}(s)G_{OL}(s)$) for DC bus voltage regulation.}
	\label{fig:anaModVal}
\end{figure}

To validate the small signal model, we consider the DC bus voltage regulation for active rectifier operation of the VSC listed in Table~\ref{TB:vscParam}. The DC bus voltage reference is set as $v^*=400$ V and $\phi=\pi/2$ is used. The control parameters are selected as $\eta = 16.63,\ \mu=0,\ R_{vir}=0.21\ \Omega (\equiv 4.9\%\ \text{p.u.}),\ \omega_c=1.2\ \text{krad/s},\ K_{pdc}=75\ \text{W/V},\ T_i=0.4s,\ \omega_z = 5\pi\ \text{rad/s}$, and $\omega_p = 30\pi\ \text{rad/s}$.
To validate the analytical model, loop-gain measurement is performed; multi-tone small signal perturbation is injected in the feedback loop and the loop gain is calculated at discrete frequency points from the recorded response \cite{multiTone}. Fig.~\ref{fig:anaModVal} shows the comparison between the analytical and measured responses.

\subsection{GFL Operation with Weak Grid}\label{sec:simGFLuwg}
\begin{figure}[htb]
	\makebox[\linewidth][c]{\includegraphics[angle = 0, clip, trim=0cm 0cm 0cm 0cm,  width=0.4\textwidth]{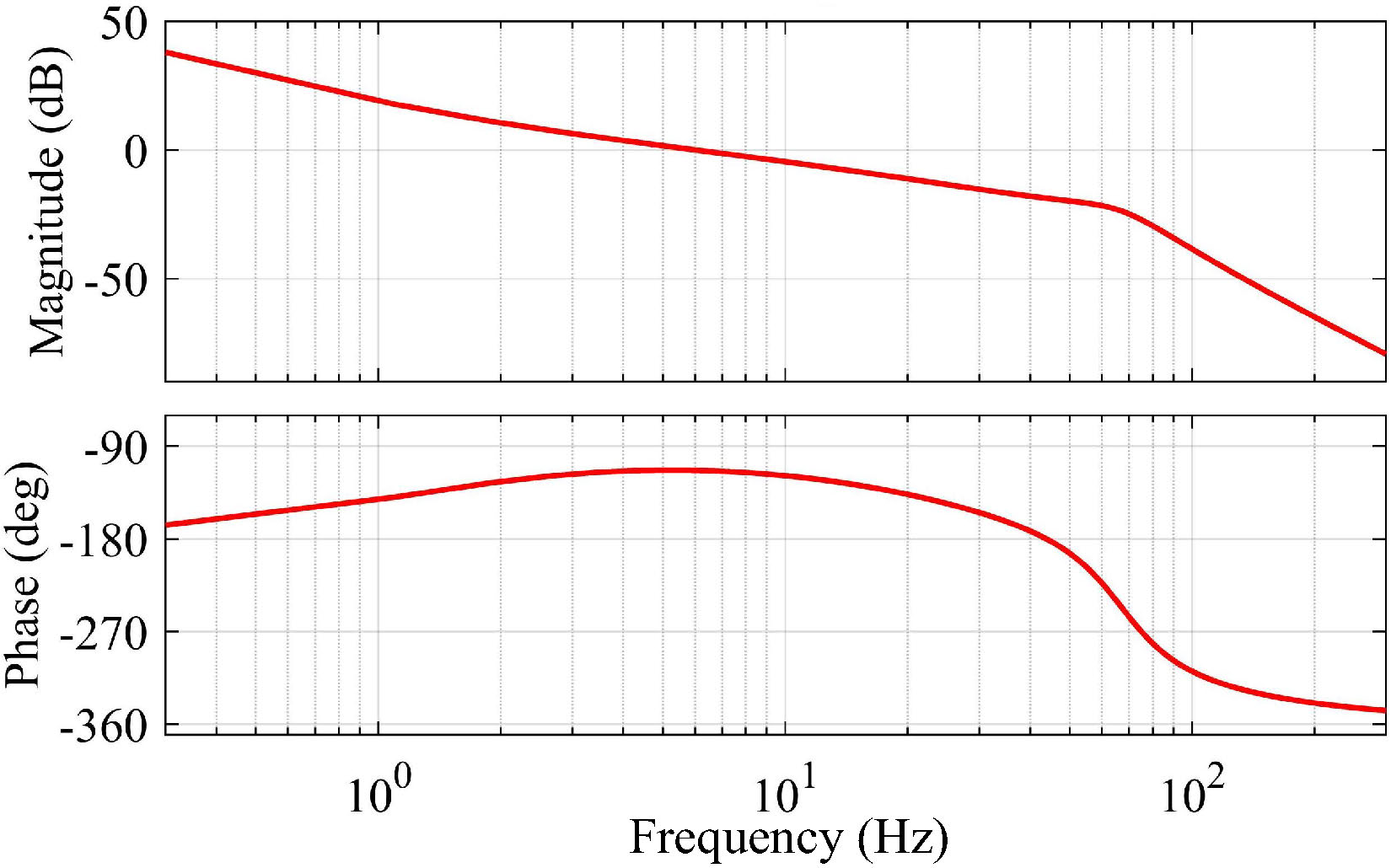}}
	\caption{Compensated open-loop response for DC bus voltage regulation with a weak grid of SCR=1.9.}
	\label{fig:uwgBode}
\end{figure}

\begin{figure}[htb]
	\makebox[\linewidth][c]{\includegraphics[angle = 0, clip, trim=0cm 0cm 0cm 0cm,  width=0.425\textwidth]{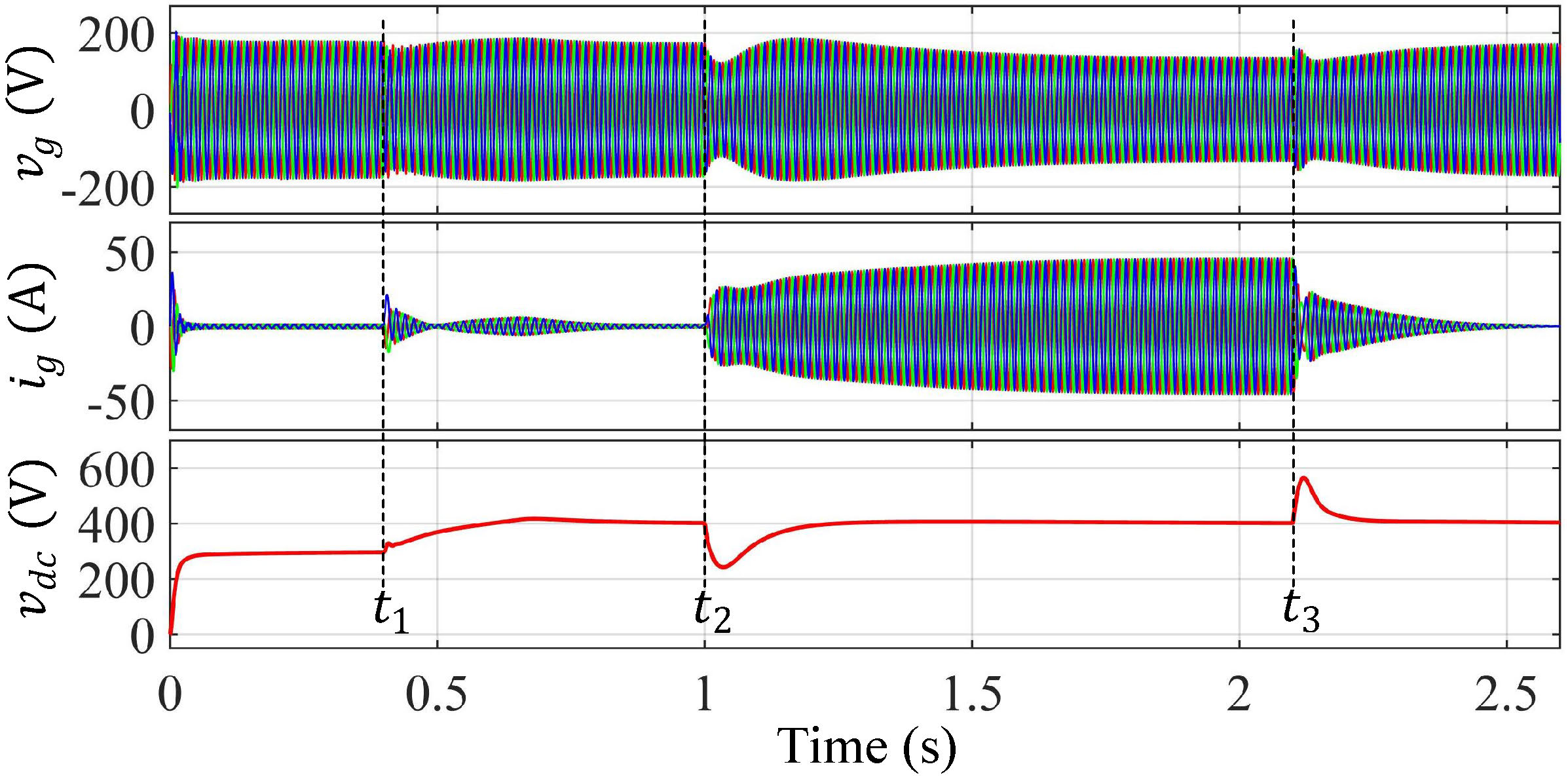}}
	\caption{Simulated response of a three-phase active rectifier for DC bus voltage regulation while connected to a weak grid of SCR=1.9.}
	\label{fig:gflUWG}
\end{figure}

To illustrate the GFL operation with weak grid, we consider the three-phase VSC described in Section~\ref{sec:anaModVal} connected to a grid with short-circuit ratio (SCR) of 1.9. Through small signal analysis, the controller is redesigned as $R_{vir}=23.6$ p.u. and $K_{pdc}=335\ \text{W/V}$ while keeping the other parameters same. The compensated open-loop response is shown in Fig.~\ref{fig:uwgBode}. Fig.~\ref{fig:gflUWG} shows the simulated response. After charging the DC bus, the closed-loop regulation is initiated at $t_1$. Subsequently, no-load to full-load and full-load to no-load transitions are introduced across the DC bus at $t_2$ and $t_3$, respectively. In both cases, the desired DC bus voltage is quickly restored.  

\subsection{Fault Ride-Through}\label{sec:simFRT}
\begin{figure}[htb]
	\makebox[\linewidth][c]{\includegraphics[angle = 0, clip, trim=0cm 0cm 0cm 0cm,  width=0.425\textwidth]{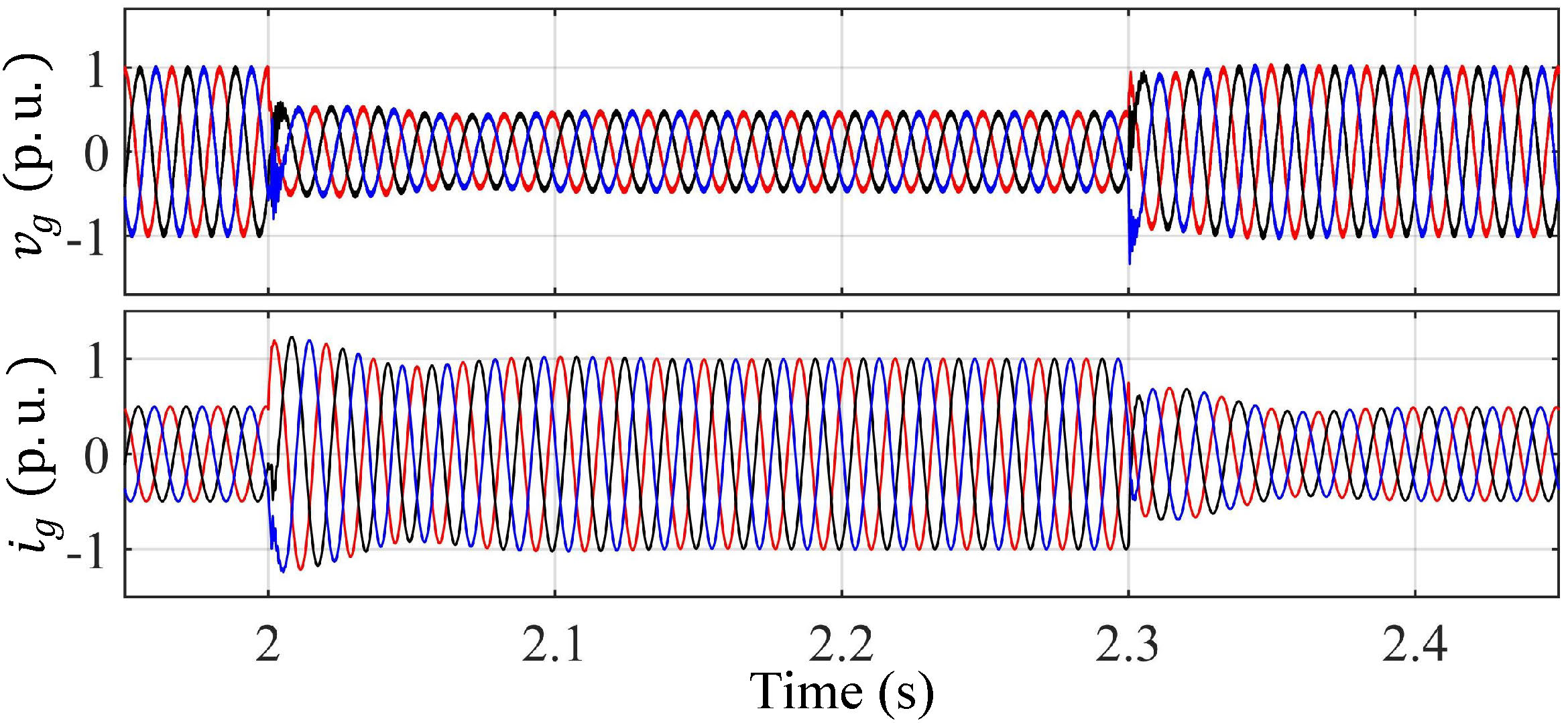}}
	\caption{Simulated converter response to a symmetrical AC fault with a voltage magnitude of 0.3 p.u. and SCR of 5.}
	\label{fig:frtStrg}
\end{figure}

\begin{figure}[htb]
	\makebox[\linewidth][c]{\includegraphics[angle = 0, clip, trim=0cm 0cm 0cm 0cm,  width=0.425\textwidth]{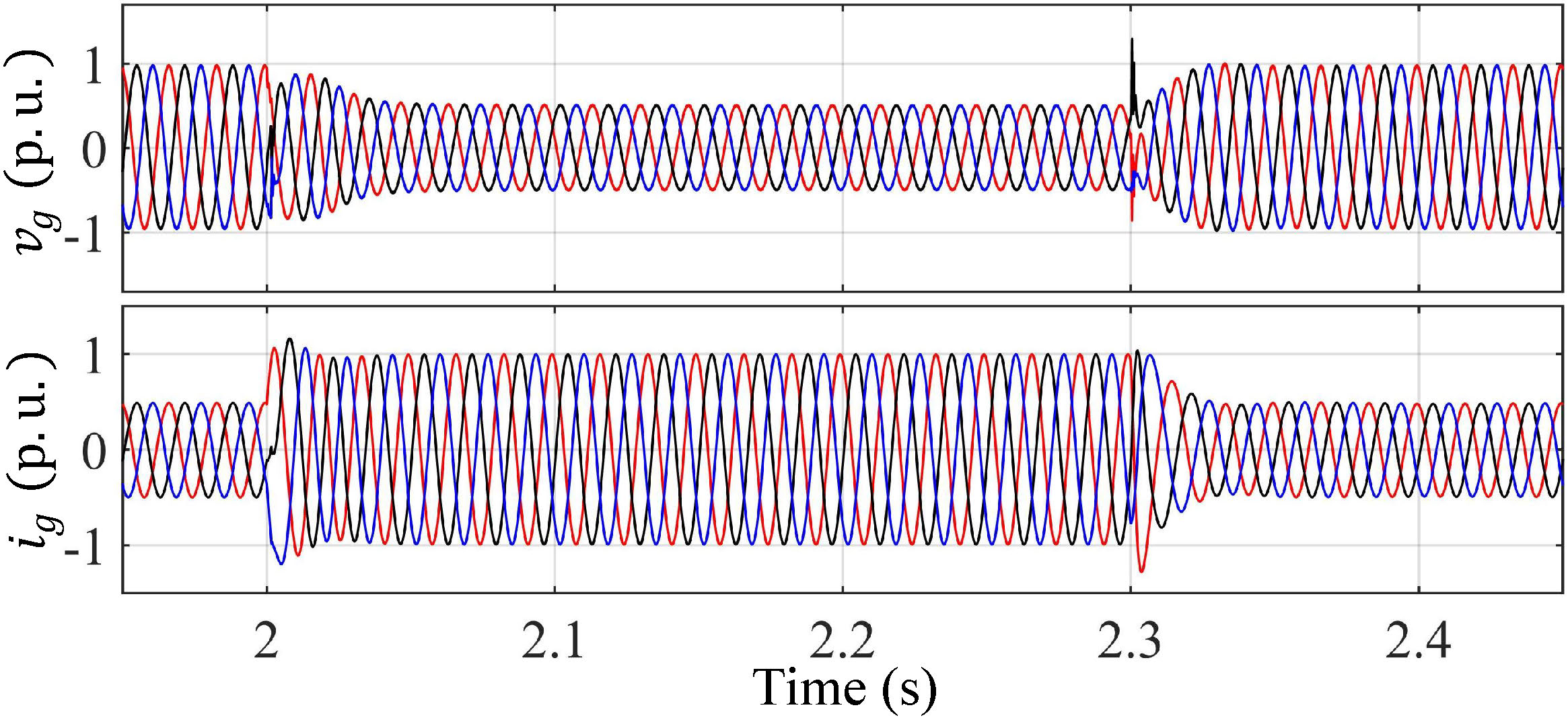}}
	\caption{Simulated converter response to a symmetrical AC fault with a voltage magnitude of 0.3 p.u. and SCR of 1.9.}	\label{fig:frtUwg}
\end{figure}

Next, the VSC is operated as a GFM converter using $\phi=\pi/2,\ \eta = 16.63,\ \mu=5.2\times 10^{-4},$ and $\ R_{vir}=0.21\ \Omega$; a voltage source is used to maintain the DC bus. The grid is modelled by an ideal AC source with an inductor to emulate different SCRs. First, we consider an SCR of 5. The control parameters are set as $V_T=0.9$ p.u., $I_T=1.1$ p.u., $I_m=1$ p.u, $R_0=5.25$ V/A, $t_f=100$ ms and $\tau_f=28$ ms, whereas the real power reference is set as $P_0=0.5$ p.u. To emulate a symmetrical AC fault, the AC source voltage magnitude is suddenly stepped down from 1 p.u. to 0.3 p.u. at $t=2$s (see Fig.~\ref{fig:frtStrg}). Once the fault is detected, the reactive power reference $Q_0$ is raised to utilize remaining current capability of the VSC as $Q_0=\sqrt{S^2_{rated}-P^2_0}$. The converter clamps the output current at 1 p.u. and once the fault is cleared, i.e. source voltage returned to 1 p.u. at $t=2.3$s, the VSC resumes normal operation. The experiment is repeated for an SCR of 1.9 and the corresponding result is shown in Fig.~\ref{fig:frtUwg}.

Note that for relatively stronger grids, to limit the initial current overshoot at the instants when a fault occurs and clears, a band-limited virtual inductance can be used in addition with the virtual resistance as $(R_{vir}+sL_{vir})/(s/\omega_c+1)$ in \eqref{eq:hcs}; $L_{vir}$ can be included in $L_e$ for small signal analysis and parameter selection described in Section~\ref{sec:ssm} and \ref{sec:dynParamSelec}. $L_{vir}=1$ mH is used for the simulation result shown in Fig.~\ref{fig:frtStrg}.   

\subsection{Validation of Droop Response}\label{sec:droopResValid}
\begin{figure}[htb]
	\makebox[\linewidth][c]{\includegraphics[angle = 0, clip, trim=0cm 0cm 0cm 0cm,  width=0.425\textwidth]{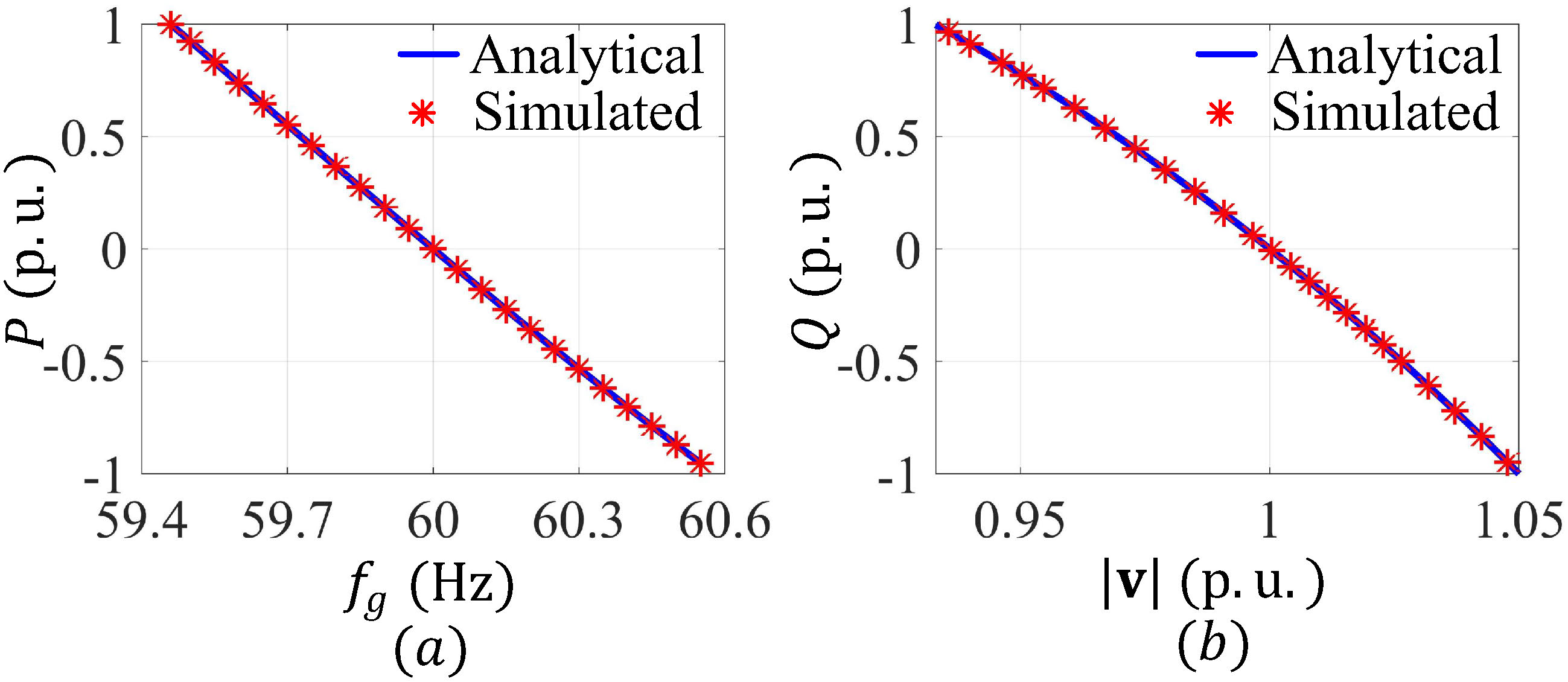}}
	\caption{Comparison of analytical and simulated droop responses - (a) real power vs. frequency for $|\mathbf{v_g}|=1$ p.u., (b) reactive power vs. voltage for $f_g=60$ Hz.}
	\label{fig:droopResValid}
\end{figure}

The VSC is run in GFM operation using $\phi=\pi/2$ and the grid frequency and voltage are varied to validate the droop response given by \eqref{eq:uocGFLpw3} and \eqref{eq:uocDRqv}. Fig.~\ref{fig:droopResValid} shows the comparison between the simulated and analytical droop responses. 

\section{Experimental Results}
The GFL and GFM operations of uVOC are validated through hardware experiments using laboratory prototypes. For all the experiments described in the following subsections, digital control is implemented using Texas Instruments' C2000 digital signal processor (DSP) TMS320F28377S.

\subsection{DC Bus Voltage Regulation (GFL Operation)}\label{sec:expResGFL}


A single-phase active rectifier with parameters listed in Table~\ref{TB:vscParam} is used. For single phase operation the rated real and reactive powers are given as $P_{rated}=3\ \text{kW}$ and $Q_{rated}=1.5\ \text{kVAR}$, respectively. To prevent propagation of measurement noise, the nominal DC bus voltage reference is used for calculating the modulating signal $\mathbf{m}$. The nominal DC bus voltage is set as $v^*=200\ \text{V}$; $\phi=\pi/2$ is used and the control parameters are selected as $\eta = 16.63,\ R_{vir}=0.21\ \Omega, K_{pdc}=40\ \text{W/V},\ T_i=0.4s,\ \omega_z = 5\pi\ \text{rad/s}$, and $\omega_p = 30\pi\ \text{rad/s}$ for a control bandwidth of $\approx 7\pi\ \text{rad/s}$ with gain margin of $\approx 25.6\ \text{dB}$ and phase margin of $\approx 71.5^o$ at no load condition. 



\begin{figure}[htb]
	\makebox[\linewidth][c]{\includegraphics[angle = 0, clip, trim=0cm 0cm 0cm 0cm,  width=0.4\textwidth]{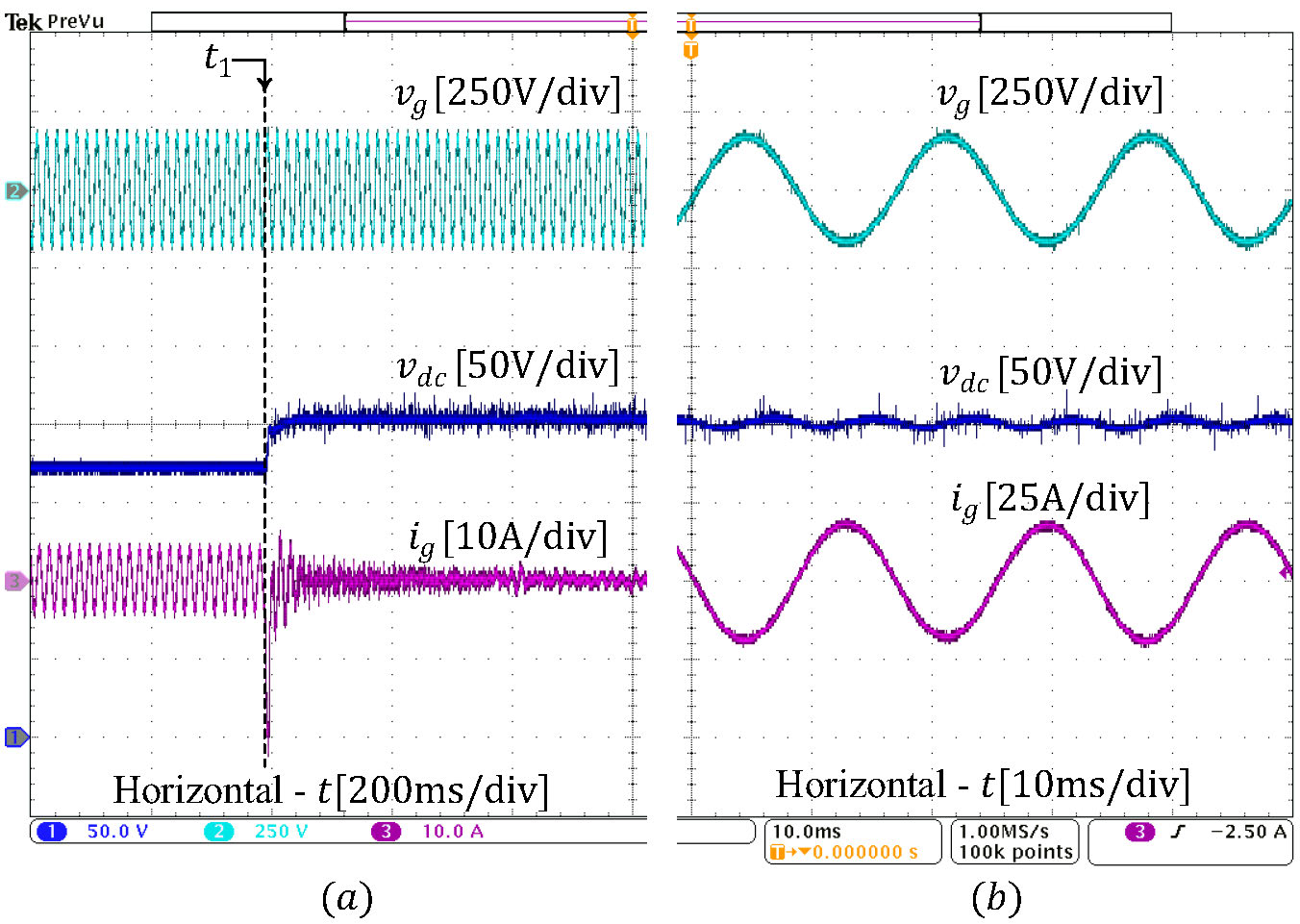}}
	\caption{DC bus voltage regulation using GFL form of uVOC - (a) start-up transient, (b) steady-state voltage and current shapes.}
	\label{fig:suGFLrec}
\end{figure}

\begin{figure}[htb]
	\makebox[\linewidth][c]{\includegraphics[angle = 0, clip, trim=0cm 0cm 0cm 0cm,  width=0.4\textwidth]{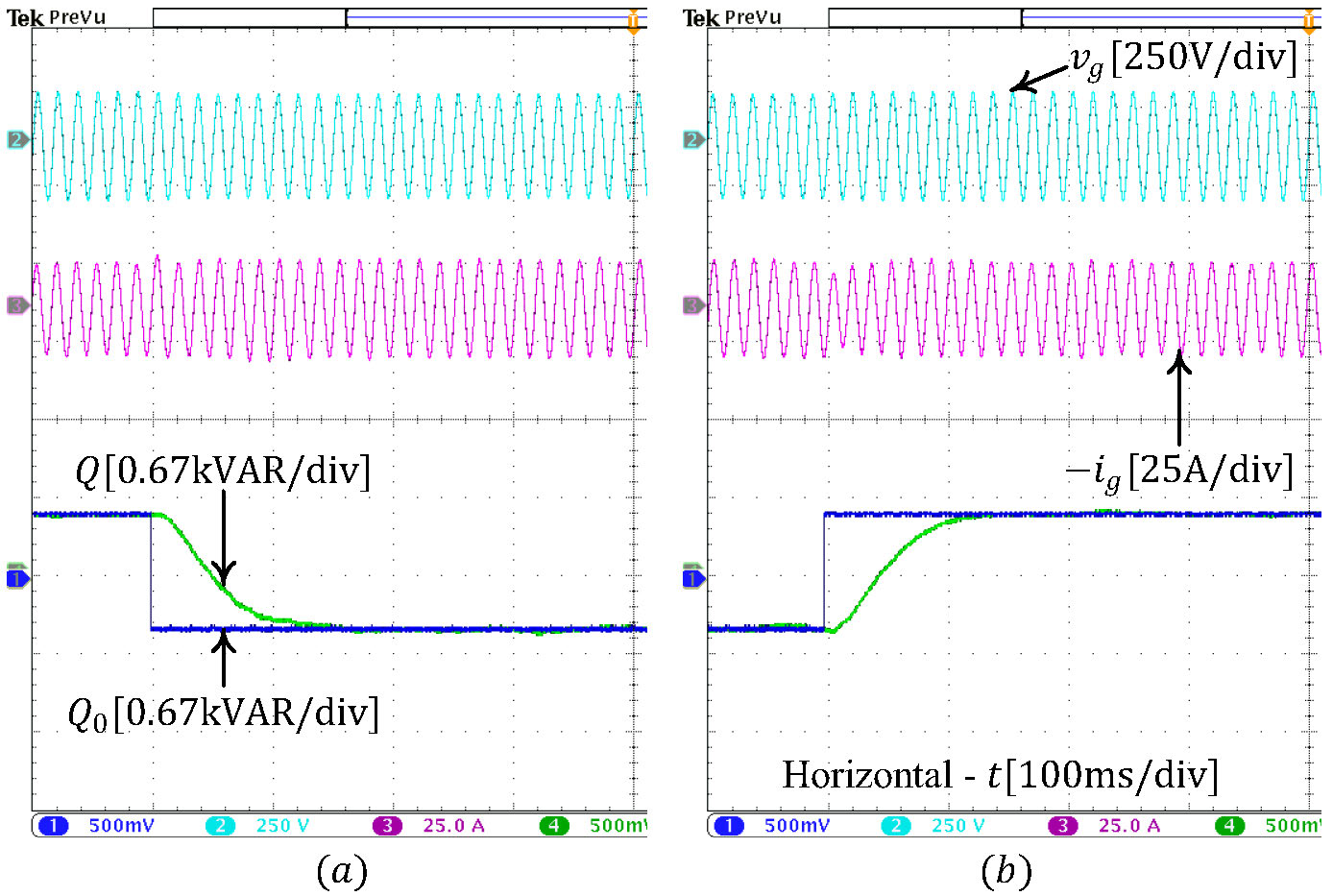}}
	\caption{Response to step change in reactive power reference- (a) $Q_0= 500\ \text{VAR}$ to $Q_0= -500\ \text{VAR}$, (a) $Q_0= -500\ \text{VAR}$ to $Q_0= 500\ \text{VAR}$.}
	\label{fig:q0Step}
\end{figure}

\begin{figure}[htb]
	\makebox[\linewidth][c]{\includegraphics[angle = 0, clip, trim=0cm 0cm 0cm 0cm,  width=0.4\textwidth]{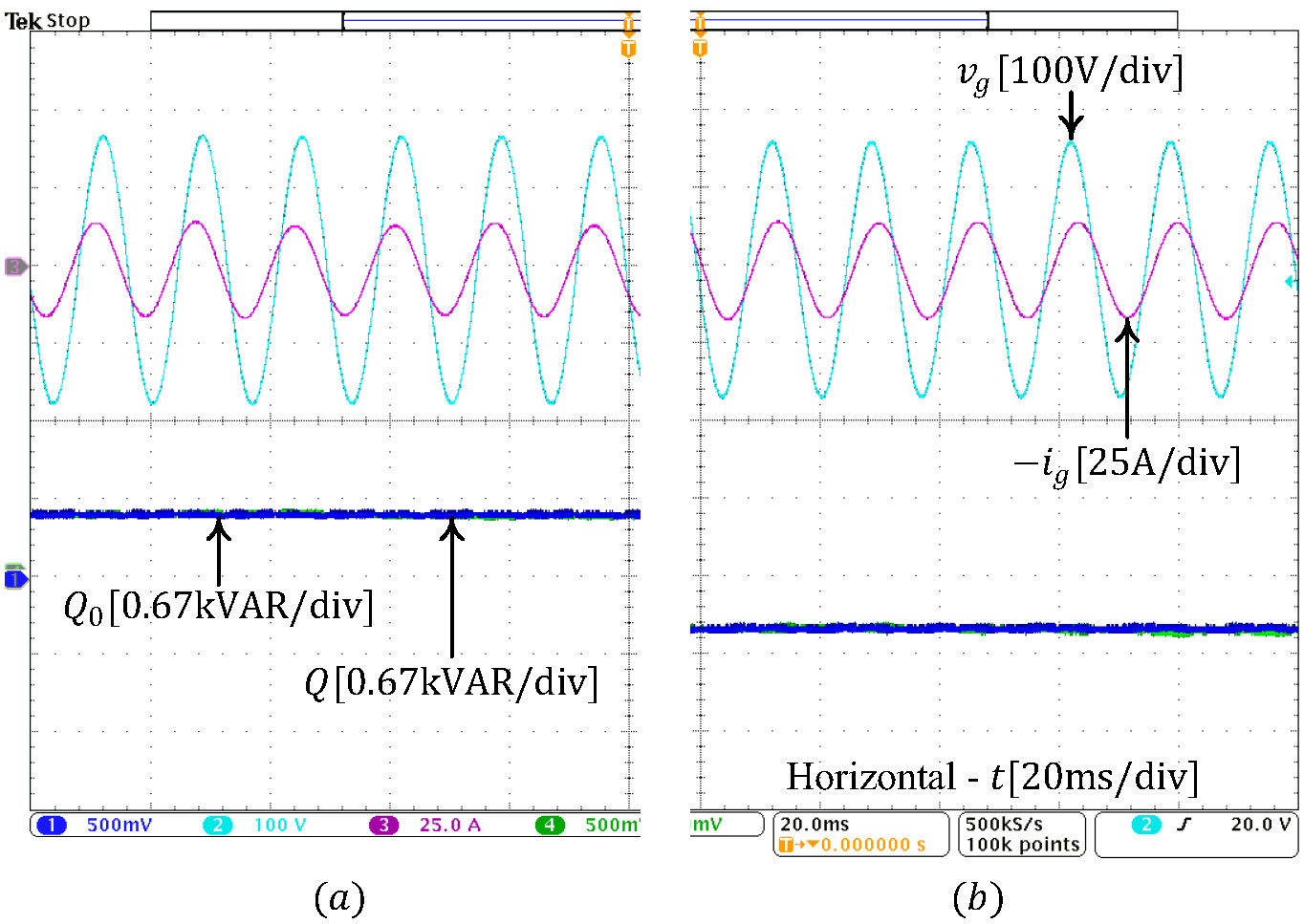}}
	\caption{Steady-state wave-shapes for - (a) $Q_0=500\ \text{VAR}$, (b) $Q_0=-500\ \text{VAR}$.}
	\label{fig:ssQ}
\end{figure}

First, an AC voltage source is used to emulate a strong grid. Fig.~\ref{fig:suGFLrec}(a) shows the start-up transient. DC bus charging and soft start-up is achieved by using the method presented in Section~\ref{sec:psGFL}; DC bus voltage regulation and converter switching are initiated at $t_1$. The DC bus voltage is quickly stabilized at the reference value. Fig.~\ref{fig:suGFLrec}(b) shows the steady-state voltage and current shapes when serving $\approx 1.5\ \text{kW}$ load and the reactive power reference is set as $Q_0=0$. Next, reactive power dispatch is introduced. The rectifier serves $\approx 1.2\ \text{kW}$ DC load, when a step change in the reactive power reference ($Q_0= 500\ \text{VAR}$ to $Q_0= -500\ \text{VAR}$) is introduced; the corresponding waveshapes are shown in Fig.~\ref{fig:q0Step}(a). The reactive power reference $Q_0$ and the output reactive power $Q$ measured by the DSP are shown on the scope using two Digital-to-Analog (DAC) channels on the DSP. The opposite step response from $Q_0= 500\ \text{VAR}$ to $Q_0= -500\ \text{VAR}$ is shown in Fig.~\ref{fig:q0Step}(b). In both cases, accurate tracking of the reactive power reference is obtained with minimal transient. The steady-state voltage and current shapes for $Q_0=-500\ \text{VAR}$ and $Q_0=500\ \text{VAR}$ are shown in Fig.~\ref{fig:ssQ}(a) and Fig.~\ref{fig:ssQ}(b), respectively.      

\begin{figure}[htb]
	\makebox[\linewidth][c]{\includegraphics[angle = 0, clip, trim=0cm 0cm 0cm 0cm,  width=0.425\textwidth]{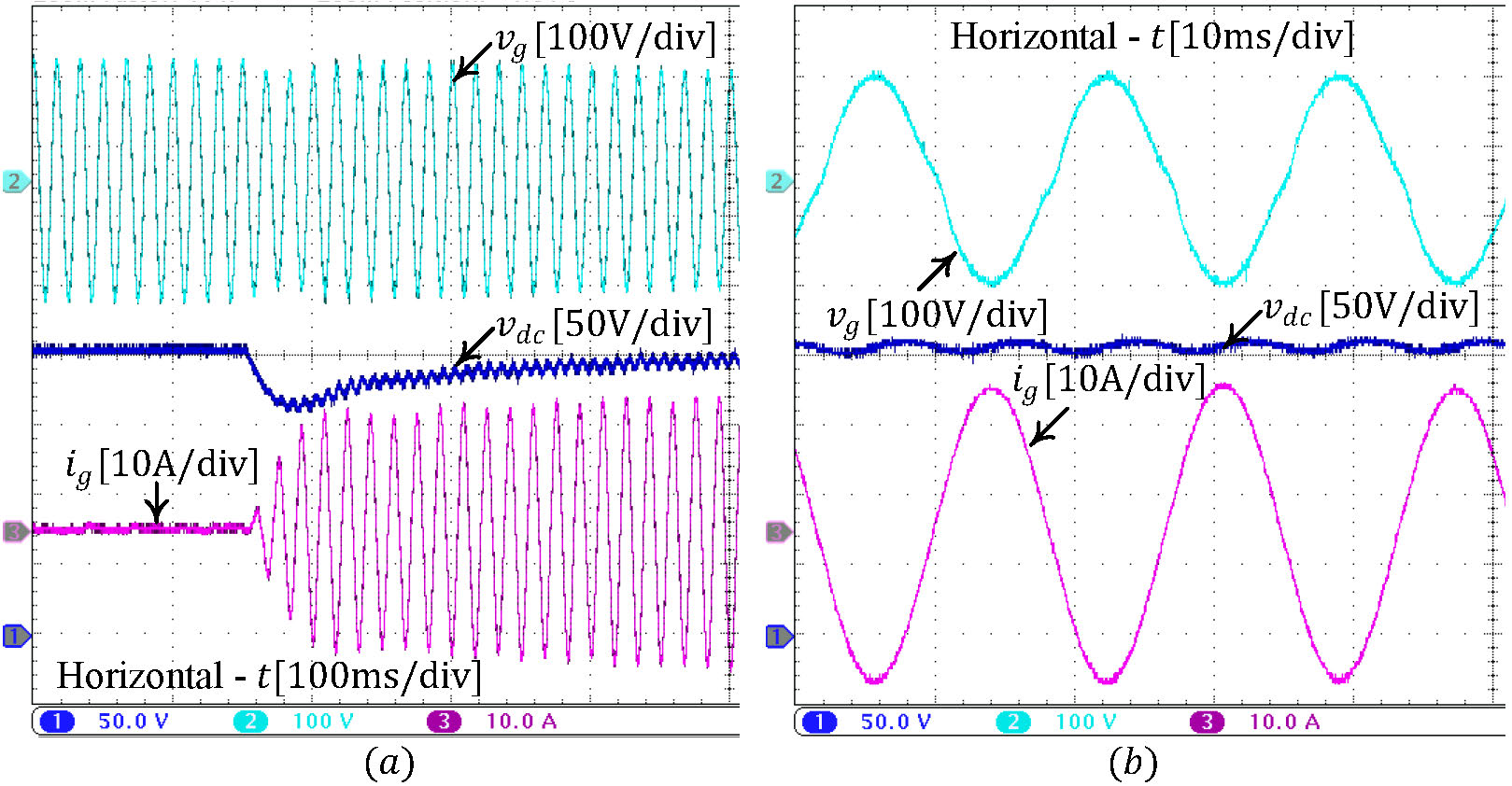}}
	\caption{GFL operation in weak grid condition (SCR=1.9) - (a) response to step change in DC load, (b) steady-state operation.}
	\label{fig:expGFLuwg}
\end{figure}

Next, to evaluate the controller under weak grid condition, $6$~mH inductor is inserted between the AC source and the VSC, which gives an SCR of $\approx 1.9$. The converter response to a step change in the DC load from $0$ to $\approx 1.4$ kW is shown in Fig.~\ref{fig:expGFLuwg}(a) and the corresponding steady-state response is shown in Fig.~\ref{fig:expGFLuwg}(b). For weak grid operation, the control parameters were redesigned as $R_{vir}=1\ \Omega, K_{pdc}=120\ \text{W/V}$, while keeping the other parameters same as that for strong grid operation. For the chosen values, the small signal model given by \eqref{eq:ssRecMode} estimates a control bandwidth of $\approx 4.4$ Hz at $P=1.4$ kW with a gain margin of $\approx 23$ dB and phase margin of $\approx 107^o$ which suggests a damped response with a settling time on the order of $1/(4.4 \text{Hz})=227$ ms, whereas during the experiment the DC bus voltage exhibits damped response and settles within $\approx 250$ ms (see Fig.~\ref{fig:expGFLuwg}(a)). It is worth noting that for this configuration the space vector oscillator achieves stable operation with an effective $X/R$ ratio of $\approx 2.8$ since $R_{vir}$ appears in the equivalent resistance seen by the oscillator; hence, assumption of a dominantly inductive grid is not required for the the SVO to retain synchronization.

\subsection{Fault Ride-Through}\label{sec:expFRT}
\begin{figure}[htb]
	\makebox[\linewidth][c]{\includegraphics[angle = 0, clip, trim=0cm 0cm 0cm 0cm,  width=0.4\textwidth]{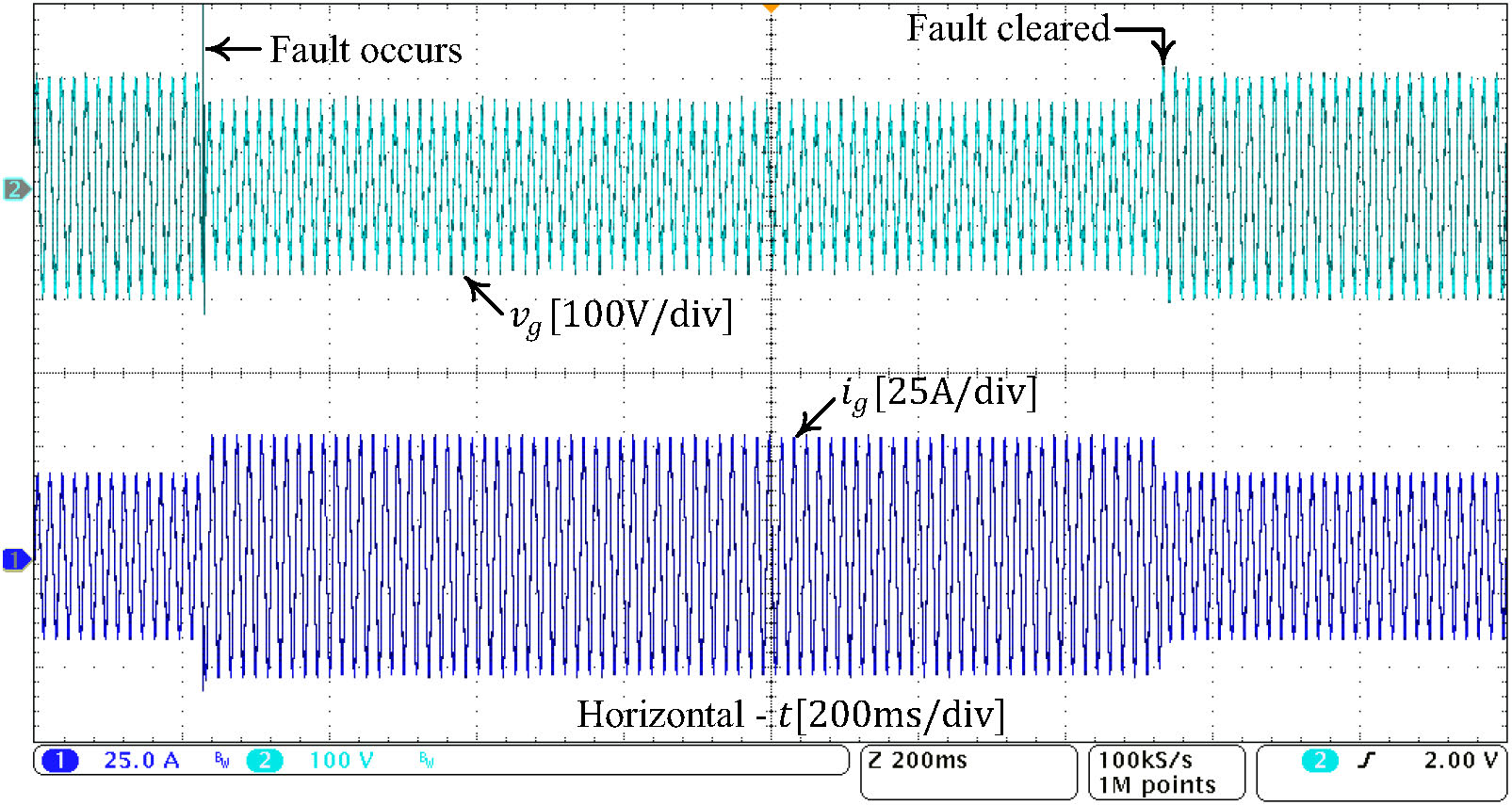}}
	\caption{Converter response to a short-circuit fault with an SCR of 1.9.}
	\label{fig:expFRTuwg}
\end{figure}

The single-phase VSC is programmed for GFM operation using the control parameters same as those used in Section~\ref{sec:simFRT}. Initially, the VSC serves a local load of $3$ kW in islanded condition; an inductor of $6$ mH is used between the load and the VSC to emulate an SCR of 1.9. A fault is introduced by shorting the load terminals and the corresponding converter response is shown in Fig.~\ref{fig:expFRTuwg}. The converter utilizes its full current capability to provide voltage support at the PoC. Once the short circuit is removed, normal operation is resumed quickly.

\begin{figure}[t]
	\makebox[\linewidth][c]{\includegraphics[angle = 0, clip, trim=0cm 0cm 0cm 0cm,  width=0.4\textwidth]{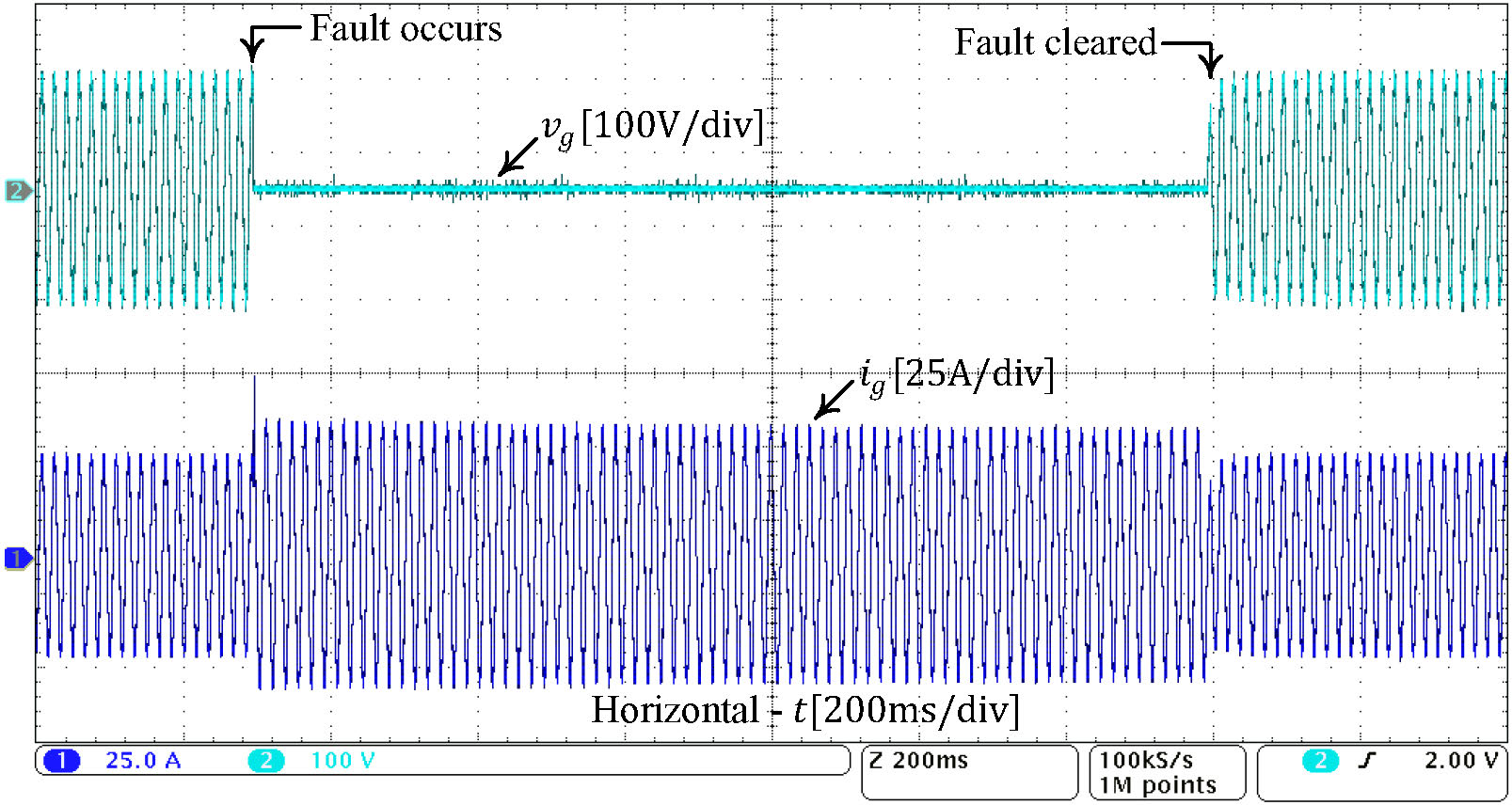}}
	\caption{Fault ride-through response when a dead-short is introduced at the converter terminal.}
	\label{fig:expFRTdeadShort}
\end{figure}

\begin{figure}[htb]
	\makebox[\linewidth][c]{\includegraphics[angle = 0, clip, trim=0cm 0cm 0cm 0cm,  width=0.425\textwidth]{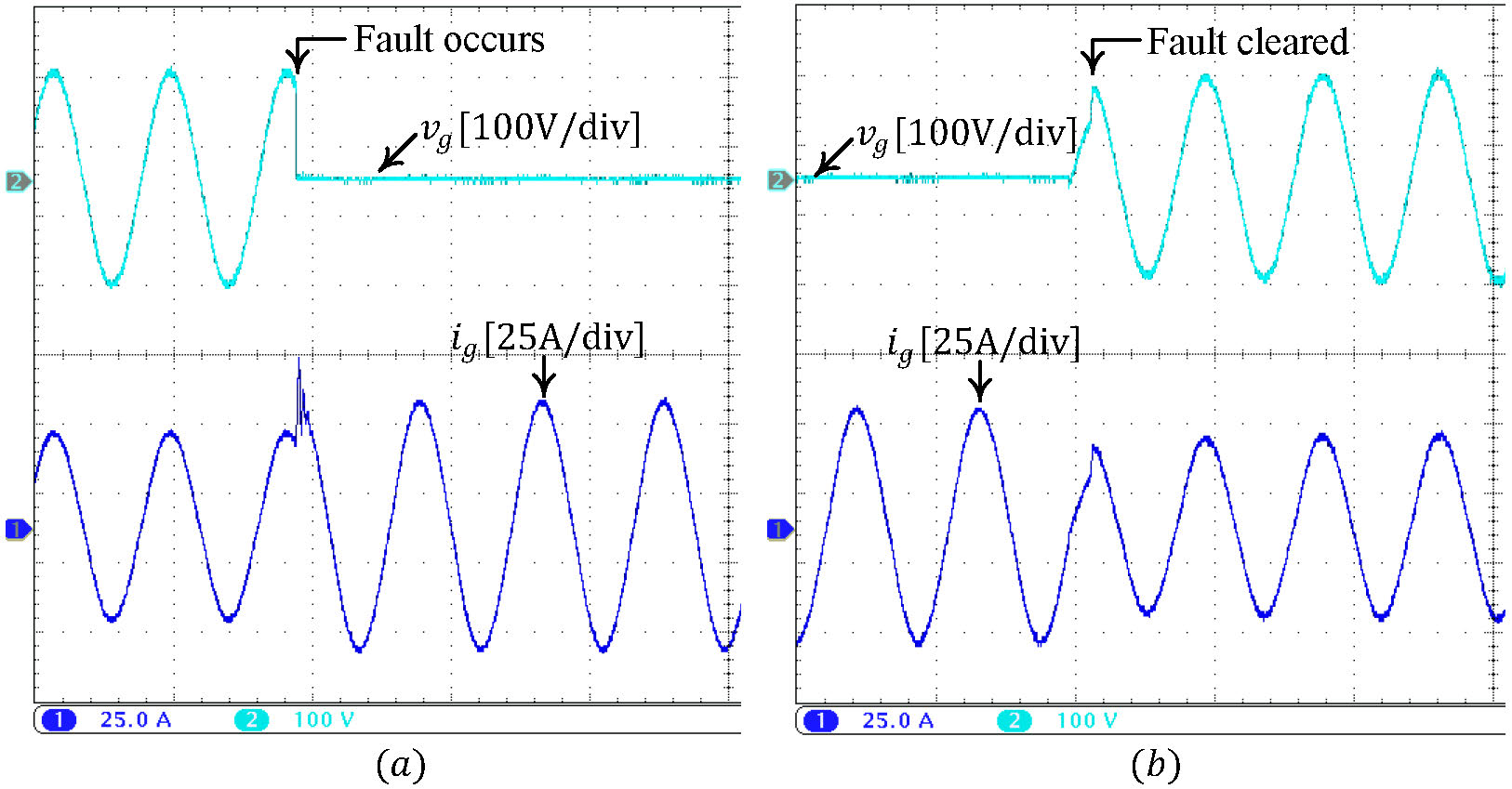}}
	\caption{Fault ride-through response to a dead-short at the converter terminal - (a) when fault occurs, (b) when fault is cleared.}
	\label{fig:expFRTzi}
\end{figure}

\noindent
Next, the inductor is removed and the load is connected at the PoC. An effective dead-short is introduced at the PoC and the response is shown in Fig.~\ref{fig:expFRTdeadShort}; the converter clamps the output current at the rated value during the fault and once the short is removed, the converter quickly returns to pre-fault condition supplying the load. The zoomed-in responses at the fault instant and the fault clearing instant are shown in Fig.~\ref{fig:expFRTzi}(a) and Fig.~\ref{fig:expFRTzi}(b), respectively. The filter capacitor $C_f$ is ignored in the small signal analysis in Sections~\ref{sec:ssm} and \ref{sec:ssmFRT}; however, the OCL compensation may trigger high frequency resonance in the range of few hundred Hz to few $kHz$. Passive damping, such as a series resistance with $C_f$, or active damping methods can be employed to avoid such resonances. An observer based active damping method similar to \cite{obadECCE} was used during the experiments, which is considered out of scope for this paper and not reported in the interest of space.

\subsection{GFM Operation}\label{sec:expGFM}
\begin{figure}[htb]
	\makebox[\linewidth][c]{\includegraphics[angle = 0, clip, trim=0cm 0cm 0cm 0cm,  width=0.475\textwidth]{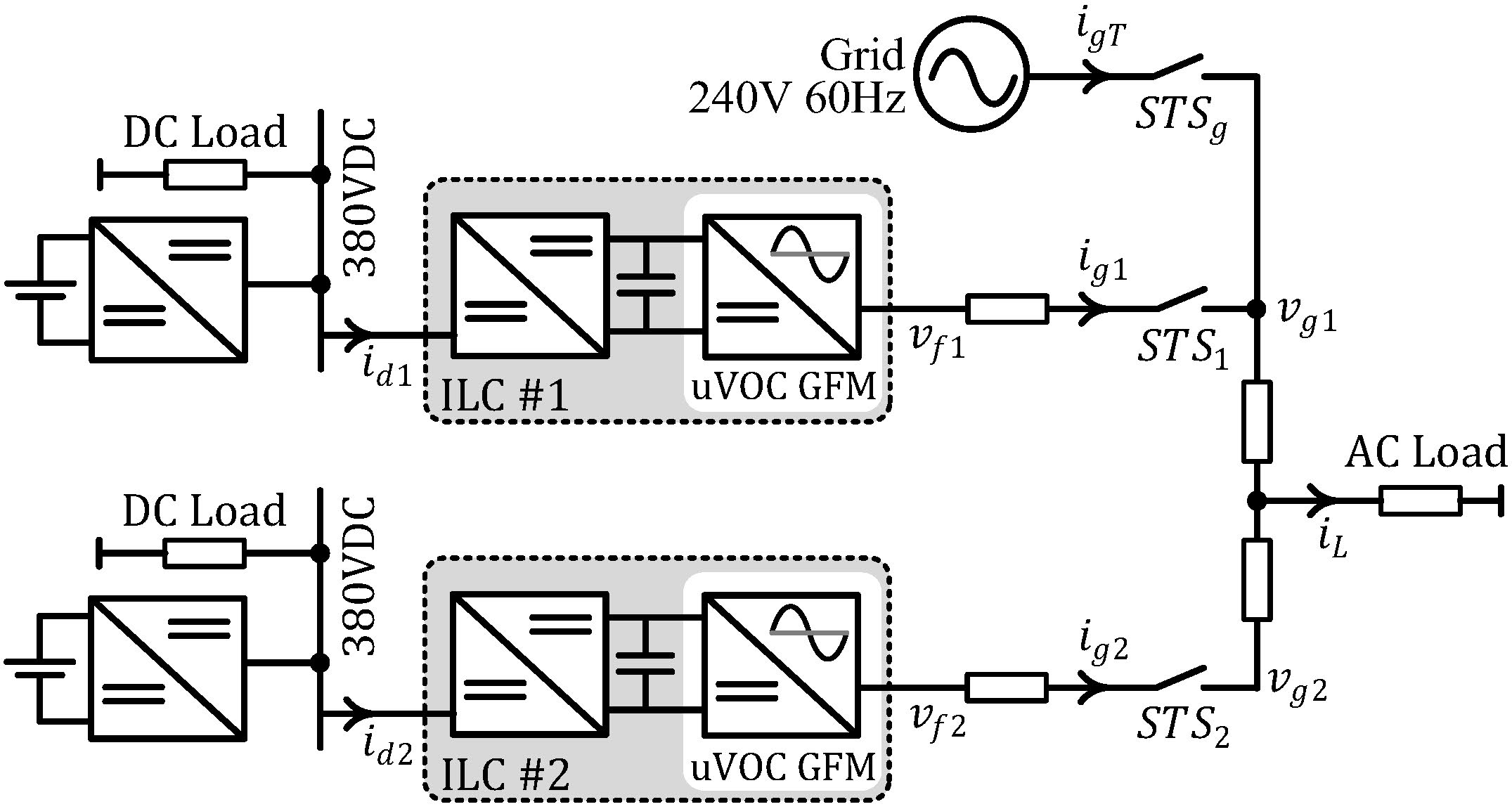}}
	\caption{Hybrid AC-DC microgrid setup.}
	\label{fig:expSetupGFM}
\end{figure}

uVOC GFM operation is validated in a hybrid AC-DC microgrid. The microgrid structure is shown in Fig.~\ref{fig:expSetupGFM}. Two battery storage units are connected to 380V DC buses through non-isolated DC/DC converters. Two interlinking converters (ILCs) are used to connect the DC systems to the AC side. Each ILC can connect/disconnect to/from the AC microgrid using static transfer switches ($STS_1$ and $STS_2$) and the AC microgrid can be islanded using another $STS_g$. The front-ends of the ILCs are rated at $P_{rated}=3\ \text{kW},\ Q_{rated}=1.5\ \text{kVAR}$, and $V_0=240.0\ \text{V}$. The control parameters are chosen as $\eta=133$, $\mu=5.3\times 10^{-4}$, and $R_{vir}=0.2\ \Omega$ for $\phi=0$ following the guidelines presented in Sections~\ref{sec:paramSelec} and $\ref{sec:dynParamSelec}$. 

\subsubsection{Islanded Operation}\label{sec:expGFMislanded}
\begin{figure}[htb]
	\makebox[\linewidth][c]{\includegraphics[angle = 0, clip, trim=0cm 0cm 0cm 0cm,  width=0.4\textwidth]{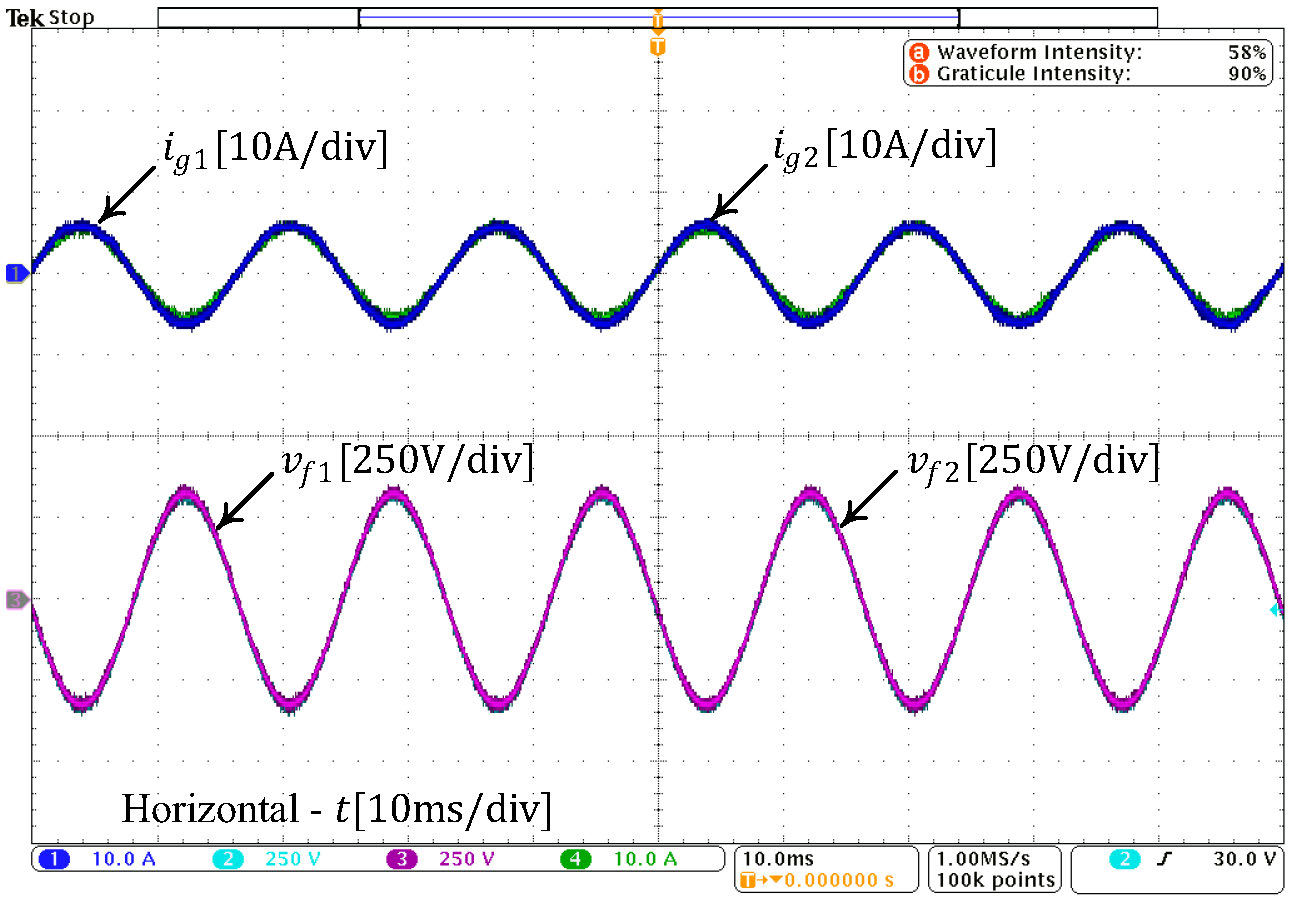}}
	\caption{ILC1 and ILC2 serving the AC load in islanded condition ($STS_g$ open; $STS_1$ and $STS_2$ are closed).}
	\label{fig:islandedGFM}
\end{figure}

\begin{figure}[htb]
	\makebox[\linewidth][c]{\includegraphics[angle = 0, clip, trim=0cm 0cm 0cm 0cm,  width=0.4\textwidth]{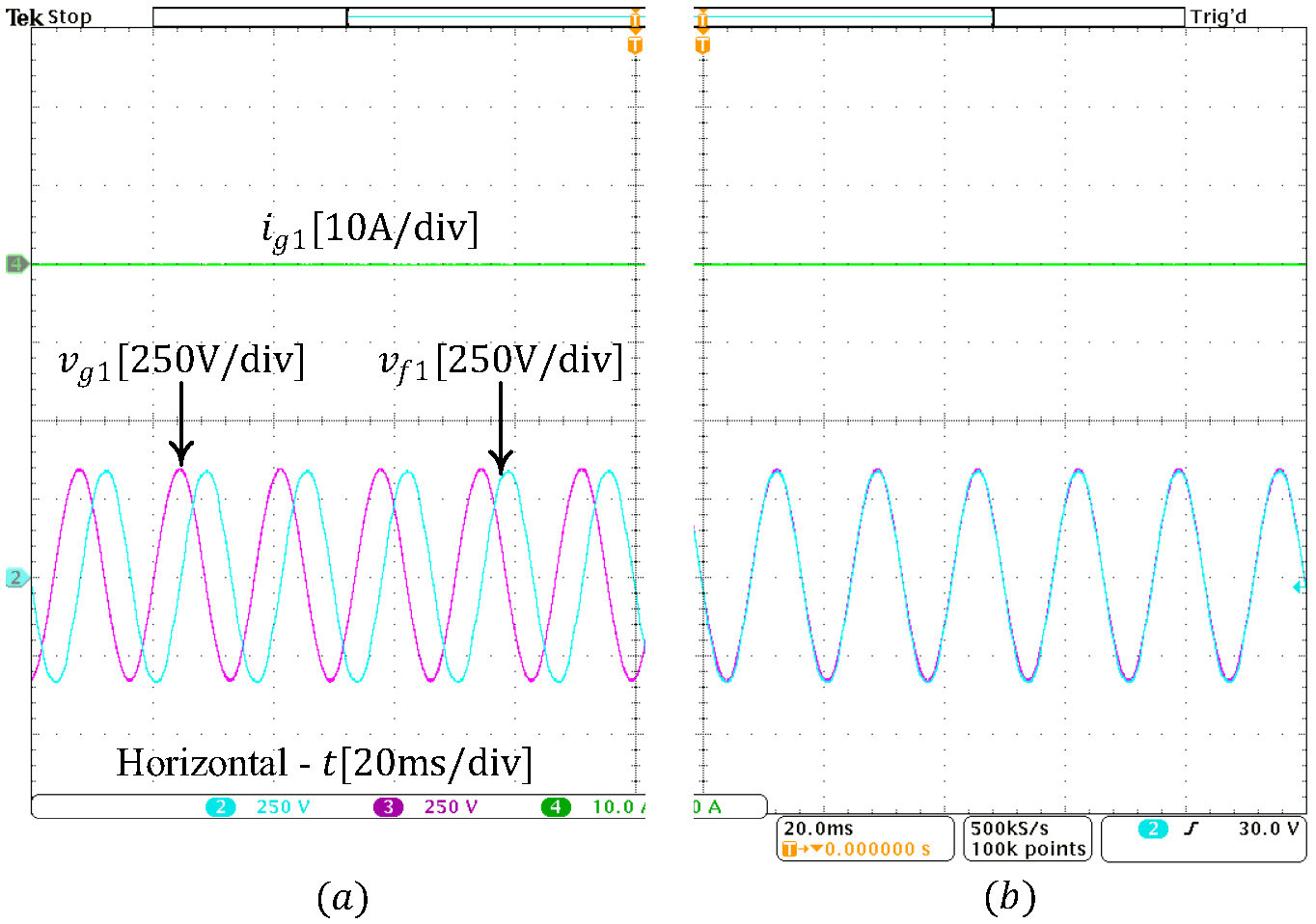}}
	\caption{Pre-synchronization of ILC1 when $STS_1$ is open and $STS_g$ is closed- (a) before synchronization, (b) synchronized condition.}
	\label{fig:preSyncGFM}
\end{figure}

\begin{figure}[htb]
	\makebox[\linewidth][c]{\includegraphics[angle = 0, clip, trim=0cm 0cm 0cm 0cm,  width=0.4\textwidth]{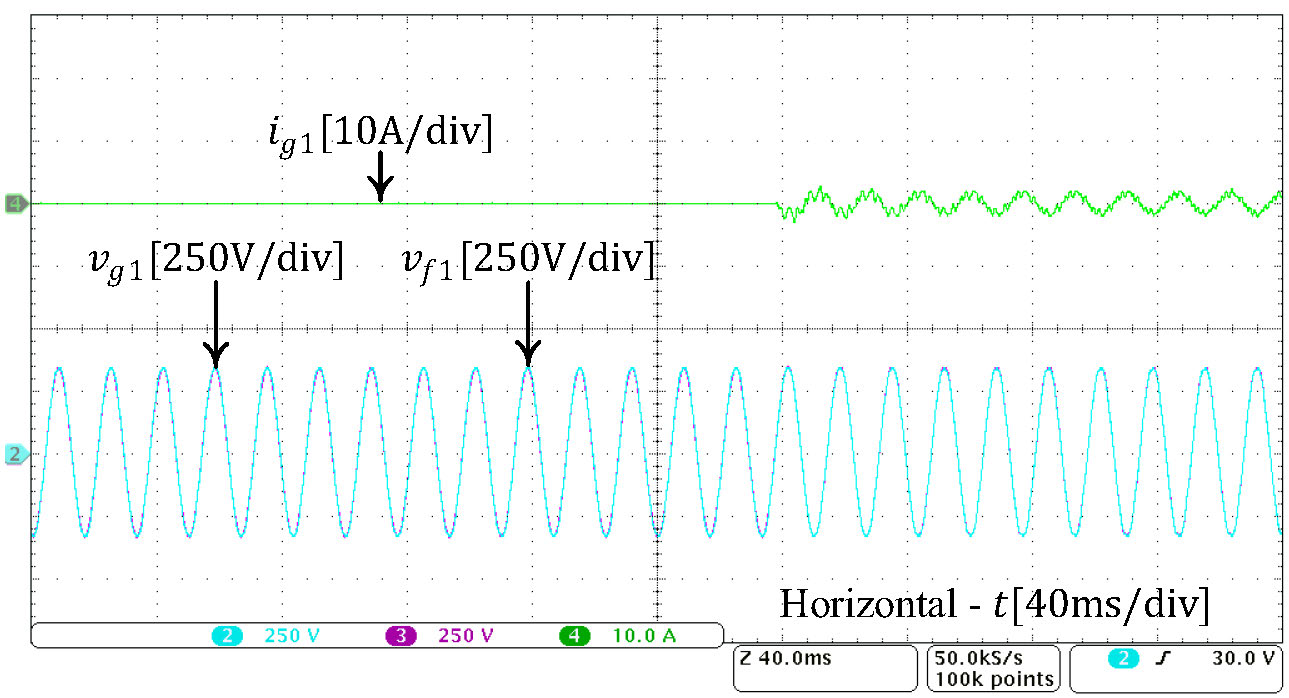}}
	\caption{ILC1 is connected to the grid by closing $STS_1$ after synchronization is achieved.}
	\label{fig:stsClosing}
\end{figure}

Islanded operation ($STS_g$ open) of the hybrid AC-DC microgrid system is illustrated in Fig.~\ref{fig:islandedGFM}, where the two interlinking converters ILC1 and ILC2 serve the AC load. Fig.~\ref{fig:preSyncGFM} shows the pre-synchronization process for ILC1, when $STS_g$ is closed and $STS_1$ is open. Before synchronization, a large phase mismatch is observed between $v_{f1}$ and $v_{g1}$ (Fig.~\ref{fig:preSyncGFM}(a)). Using the proposed pre-synchronization method, synchronization is achieved, shown in Fig.~\ref{fig:preSyncGFM}(b). The speed of the pre-synchronization process varies as a function of the initial phase, frequency, and voltage magnitude difference; trajectory analysis or numerical simulation may be used to determine the expected longest time required under various initial conditions. During the experiment shown in Fig.~\ref{fig:preSyncGFM}, pre-synchronization is achieved within $<1$ s. At this condition, ILC1 can be connected to the grid by closing $STS_1$. The corresponding transient is shown in Fig.~\ref{fig:stsClosing}.

\begin{figure}[htb]
	\makebox[\linewidth][c]{\includegraphics[angle = 0, clip, trim=0cm 0cm 0cm 0cm,  width=0.4\textwidth]{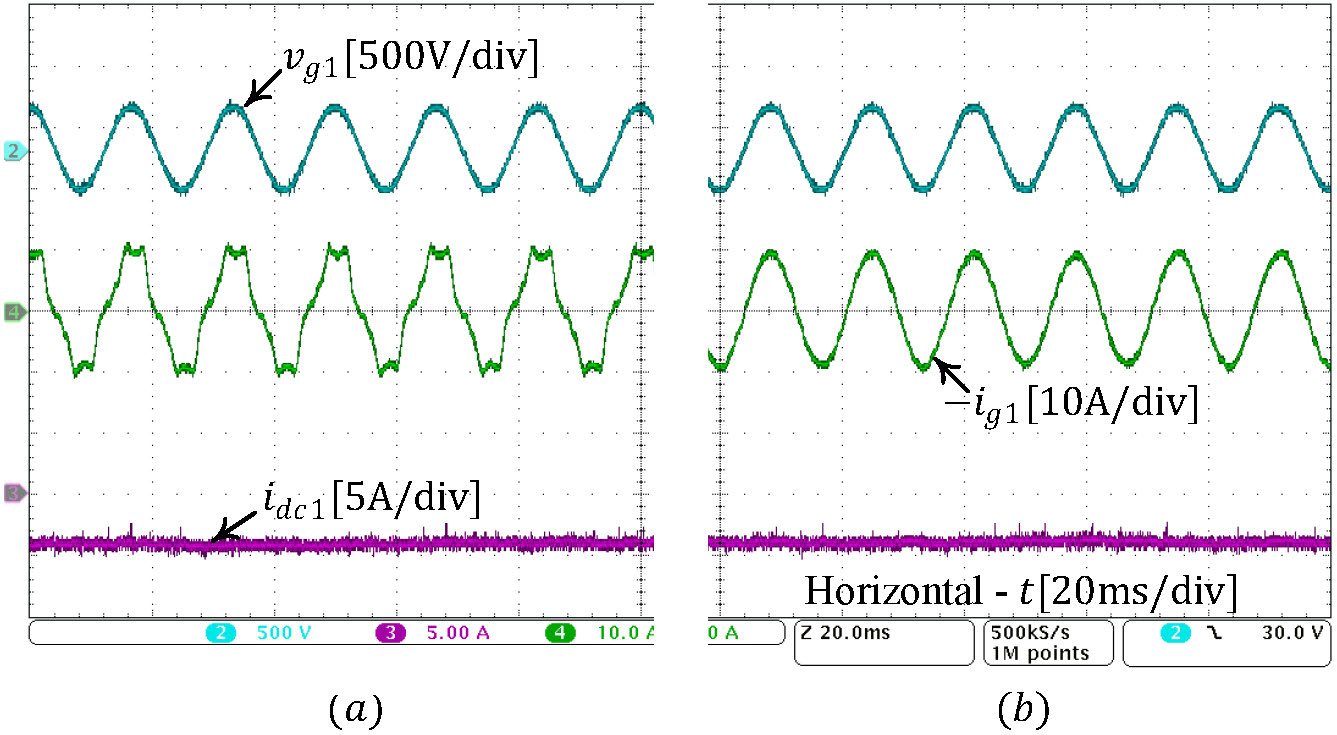}}
	\caption{Harmonic compensation in converter output current- (a) without harmonic suppression filters, (b) with harmonic compensation using \eqref{eq:hcs}.}
	\label{fig:wowoHcs}
\end{figure}

\begin{figure}[htb]
	\makebox[\linewidth][c]{\includegraphics[angle = 0, clip, trim=0cm 0cm 0cm 0cm,  width=0.4\textwidth]{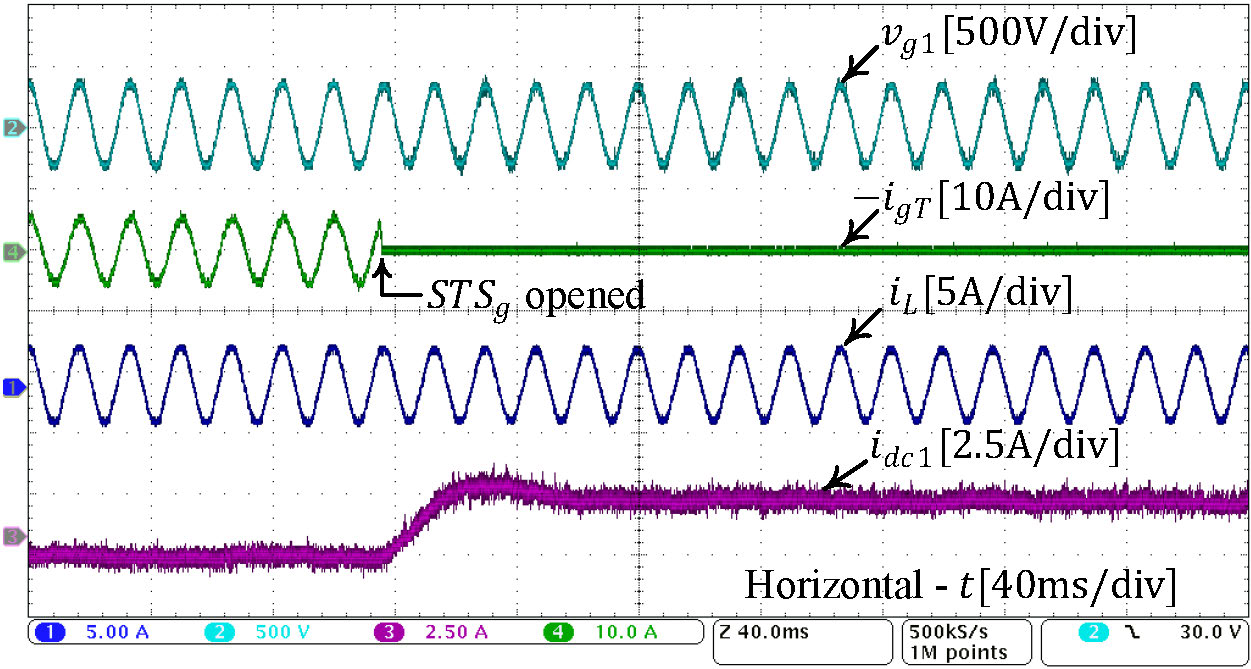}}
	\caption{AC load does not experience any disturbance (uninterrupted load current $i_L$) when the system is islanded unintentionally by opening $STS_g$.}
	\label{fig:unIntIslanding}
\end{figure}



\subsubsection{Grid-Connected Operation}\label{sec:expGT}
In grid-connected operation, the converter output current without using the harmonic compensation method is illustrated in Fig.~\ref{fig:wowoHcs}(a). Evidently, higher order harmonics are present in the output current. Using resonant filters given in \eqref{eq:hcs} at $h=3, 5, 7, 9, 11,$ and $13$, the distortion is compensated and the corresponding current shape is shown in Fig.~\ref{fig:wowoHcs}(b). 
Next, unintentional islanding is demonstrated in Fig.~\ref{fig:unIntIslanding}. Prior to disconnecting from the grid by opening $STS_g$, the ILC1 draws real power from the grid ($i_{dc1}<0$). At the opening of $STS_g$, marked by the transition in grid current $i_{gT}$, the power-flow through the ILC1 reverses immediately and the load current $i_L$ does not experience any noticeable disturbance. 

Oscillator based control, such as dVOC, was initially conceived based on a number of simplifying assumptions; all sources in the network were assumed to be oscillator based and the power converters were treated as ideal voltage sources ignoring the nonidealities of real systems such as harmonic distortion originating from network/grid side or deadtime/blanking time of the power devices. However, such simplifying assumptions are not required for the analysis and design of uVOC which is validated through the experimental results showing grid-tied operation for both GFL and GFM applications and harmonic current suppression. It is worth noting that uVOC is still an emerging technology in its early developmental phase. Established methods such as VSM enables synthetic inertia emulation and decades of research has established standard practices and implementation guidelines under a variety of application conditions. The established controllers offer intuitive insight into their synchronization mechanism leading to simplified design. In this work, physical interpretation of the synchronization process of uVOC is presented. Furthermore, effective fault ride-through capability and operation under weak grid conditions have been demonstrated. Overall, the key distinctions of uVOC in context of the state-of-the art GFM controllers are summarized in Table~\ref{tb:comp}.  

\renewcommand{\arraystretch}{1}
\begin{table}[h]
\vspace{-10pt}
\centering
\caption{Comparison among Controllers}
\begin{tabular}{p{4.5cm}p{0.65cm}p{0.4cm}p{0.55cm}p{0.55cm}}
\toprule
& Droop/ VSM & PSC & dVOC & uVOC\\
\midrule
GFL operation with weak grid & N/A & Yes & No & Yes\\
Controller switch needed during fault & Yes & Yes & N/A & No\\
Inertia emulation capability & Yes & No & No & No\\
AC voltage control loop needed & Yes & No & No & No\\

\bottomrule
\end{tabular}
\label{tb:comp}
\end{table}

\section{Conclusion}
An oscillator based control, namely uVOC, is presented for grid connected and islanded voltage source converters taking advantage of the rigorous analytical framework provided by such oscillator based methods. The proposed uVOC provides a comprehensive analysis and unified design solution for grid following and grid forming converters. The presented analysis provides an intuitive physical interpretation of the synchronization mechanism of uVOC. Through experiments, the proposed GFL controller is shown to retain synchronization with strong and weak grids which enables to avoid synchronization issues of PLLs under weak grid operation. DC bus voltage regulation is also demonstrated using the proposed GFL controller. The GFM controller achieves seamless transition from islanded to grid-connected mode using the PLL-less pre-synchronization method as well as serves local loads without interruption in an event of unintentional islanding. Enhanced fault-ride through capability is demonstrated without the need for switching to a back-up controller. 

\appendices
\section{Droop Response of Oscillator}\label{apndx:a}


For an arbitrary rotation angle $\phi$, the SVO GFM dynamics can be derived as

\begin{equation}\label{eq:uocGFMarbitPhi}
  \begin{split}
    \frac{d}{dt}(\mathbf{v}) &= [j\left\{\omega_0+\eta (e_{iP} \sin{\phi} + e_{iQ} \cos{\phi})\right\}\\
    &+ \left\{\mu (V^2_{p0}-V^2_p) + \eta (e_{iP} \cos{\phi} - e_{iQ} \sin{\phi})\right\}]\mathbf{v}.
  \end{split}
\end{equation}

\noindent
Comparing \eqref{eq:derSV} and \eqref{eq:uocGFMarbitPhi}, the dynamics along the $d$ axis can be found as

{\small
\begin{equation}\label{eq:dAxisDyn}
  \begin{split}
  \Dot{V} &= 2\mu V (V^2_0-V^2) + \frac{\eta}{NV}\left[(P_0-P)\cos{\phi}+(Q_0-Q)\sin{\phi}\right].
  \end{split}
\end{equation}
}
\normalsize

\noindent
Here, $V_{0}=V_{p0}/\sqrt{2}$ and $V=V_p/\sqrt{2}$. By setting $\frac{d}{dt}(V)=0$ and from the dynamics along the $q$-axis, the instantaneous droop response can be derived as follows: 

{\small
\begin{equation}\label{eq:uocVdroop}
  \begin{split}
    V^2 &=  V^2_0 + \frac{\eta}{2 \mu NV^2}\left[(P_0-P)\cos{\phi}+(Q_0-Q)\sin{\phi}\right];
  \end{split}
\end{equation}

\begin{equation}\label{eq:qAxisDyn}
  \begin{split}
  \omega = \omega_0 + \frac{\eta}{NV^2}\left[(P_0-P)\sin{\phi} - (Q_0-Q)\cos{\phi}\right].
  \end{split}
\end{equation}
}
\normalsize
\noindent
The operating voltage can be derived by solving \eqref{eq:uocVdroop} as

\small
\begin{equation}\label{eq:opV}
  \begin{split}
  V = \frac{V_0}{\sqrt{2}}\left[1+\left[1+\frac{2\eta\left[(P_0-P)\cos{\phi}+(Q_0-Q)\sin{\phi}\right]}{\mu N V^4_0}\right]^{\frac{1}{2}}\right]^{\frac{1}{2}}.
  \end{split}
\end{equation}
\normalsize

\section{Design of $\eta$ and $\mu$}\label{apndx:b}
First, we consider $\phi=\pi/2$ which relates real power with instantaneous frequency and reactive power with instantaneous voltage vector magnitude.

\subsubsection{$P-\omega$, $Q-V$ Droop ($\phi=\pi/2$)} 
For $\phi=\pi/2$, the operating point is given by

\begin{equation}\label{eq:pwDroop1}
  \begin{split}
  V &= \frac{V_0}{\sqrt{2}}\left[1+\left[1+\frac{2\eta}{N\mu V^4_0}(Q_0-Q)\right]^{\frac{1}{2}}\right]^{\frac{1}{2}};
  \end{split}
\end{equation}
\begin{equation}\label{eq:pwDroop2}
  \begin{split}
  \omega &= \omega_0 + \frac{\eta}{NV^2}(P_0-P).
  \end{split}
\end{equation}

\noindent
For the selection of $\mu$ and $\eta$, we set $P_0=0,\ Q_0=0$ which results in $V=V_0,\ \omega=\omega_0$ for $P=0,\ Q=0$. The real power output should reach the rated value at the maximum allowable frequency deviation as

\begin{equation}\label{eq:wmax}
  \begin{split}
  P_{rated}=\frac{NV^2_{max}}{\eta}\Delta \omega_{max},
  \end{split}
\end{equation}

\noindent
where, $V_{max}$ denotes the maximum voltage RMS. $\eta$ can be selected using \eqref{eq:wmax}. The reactive power output for a given voltage $V$ can be obtained from \eqref{eq:pwDroop1} as  

\begin{equation}\label{eq:QasFofV}
  \begin{split}
  Q=\frac{N\mu}{2 \eta}\left\{V^4_0-(2V^2-V^2_0)^2\right\}.
  \end{split}
\end{equation}

\noindent
For a given $\Delta V_{max}$, $|Q|_{V=V_{max}}>|Q|_{V=V_{min}}$ and therefore, to ensure $|Q|\leq Q_{rated}$ over the entire operating range, $\mu$ is selected using 

\begin{equation}\label{eq:etaMu}
  \begin{split}
  \frac{\mu}{\eta}=\frac{2Q_{rated}}{N}\frac{1}{(2V_{max}^2-V^2_0)^2-V^4_0}.
  \end{split}
\end{equation}

\subsubsection{$P-V$, $Q-\omega$ Droop ($\phi=0$)} For $\phi=0$, similar analysis can be done to obtain

\begin{equation}\label{eq:wmax1}
  \begin{split}
  Q_{rated}=\frac{NV^2_{max}}{\eta}\Delta \omega_{max};
  \end{split}
\end{equation}
  
\begin{equation}\label{eq:etaMu1}
  \begin{split}
  \frac{\mu}{\eta}=\frac{2P_{rated}}{N}\frac{1}{(2V_{max}^2-V^2_0)^2-V^4_0}.
  \end{split}
\end{equation}

\section{Small Signal Model of uVOC Based VSC}\label{apndx:d}
Using \eqref{eq:polOscSRF}, \eqref{eq:iDyn}, \eqref{eq:PQSRF}, and \eqref{eq:dcBusDyn}, the dynamics of the uVOC based VSC can be organized as
  
\begin{equation}\label{eq:cSys}
  \begin{split}
  \Dot{x}_c = \mathrm{F}(x_c,u).
  \end{split}
\end{equation}
\noindent
Here, $x_c=[\Delta I_d\ \Delta I_q\ \Delta V\ \Delta \theta_s\ \Delta v_{dc}]^T$ and $u=[\Delta P_0\ \Delta Q_0]^T$. The linearized system is obtained as

\begin{equation}\label{eq:lSys}
  \begin{split}
  \Dot{x}_c = A x_c &+ B u;\\
  A = \left[\frac{\partial \mathrm{F}}{\partial x_c}\right]_{(x_{eq},u_{eq})} &; \quad B = \left[ \frac{\partial \mathrm{F}}{\partial u}\right]_{(x_{eq},u_{eq})}.
  \end{split}
\end{equation}

\noindent
For ease of analysis for different forms of the controller, the state vector $x_c$ is partitioned as $x_c=[x^T\ v_{dc}]^T$ and the linearized matrices $A$ and $B$ are partitioned accordingly to obtain \eqref{eq:ssMod}. The detail forms of the different matrices in \eqref{eq:ssMod} are given in \eqref{eq:ssMatrices}.

\renewcommand{\arraystretch}{1.5}
\begin{figure*}[t]
\begin{equation}\label{eq:ssMatrices}
  \begin{split}
   &\left[{\begin{array}{c|c}
          A_{11} & A_{12}\\
          \hline
          A_{21} & A_{22}\\
    \end{array}}\right]_{\phi=\frac{\pi}{2}}=
    \left[{\begin{array}{cccc|c}
         -\frac{R_e}{L_e} & \omega_* & \frac{k_v}{L_e}\cos{(\theta_s)} & -\frac{k_v V}{L_e}\sin{(\theta_s)} & \frac{V}{V^*_{dc} L_e}\cos{(\theta_s)}\\
         -\omega_* & -\frac{R_e}{L_e} & \frac{k_v}{L_e}\sin{(\theta_s)} & \frac{k_v V}{L_e}\cos{(\theta_s)} & \frac{V}{V^*_{dc} L_e}\sin{(\theta_s)}\\
         -\eta k_v \sin{(\theta_s)} & \eta k_v \cos{(\theta_s)} & 2\mu(V^2_0-3V^2)-\frac{\eta Q_0}{NV^2} & -\eta k_v \xi_1 & -\eta \xi_2/(V^*_{dc})\\
         -\frac{\eta k_v}{V} \cos{(\theta_s)} & -\frac{\eta k_v}{V} \sin{(\theta_s)} & -\frac{2\eta P_0}{NV^3}+\frac{\eta k_v \xi_1}{V^2} & \eta k_v \xi_2/V & -\eta \xi_1/(V V^*_{dc})\\
         \hline
         -\frac{N V}{C_{dc}V^*_{dc}}\cos{(\theta_s)} & -\frac{N V}{C_{dc}V^*_{dc}} \sin{(\theta_s)} & -N \xi_1/(C_{dc}V^*_{dc}) & N V \xi_2/(C_{dc}V^*_{dc}) & 0\\ 
    \end{array}}\right];\\
    &\left[{\begin{array}{c|c}
          A_{11} & A_{12}\\
          \hline
          A_{21} & A_{22}\\
    \end{array}}\right]_{\phi=0}=
    \left[{\begin{array}{cccc|c}
         -\frac{R_e}{L_e} & \omega_* & \frac{k_v}{L_e}\cos{(\theta_s)} & -\frac{k_v V}{L_e}\sin{(\theta_s)} & \frac{V}{V^*_{dc} L_e}\cos{(\theta_s)}\\
         -\omega_* & -\frac{R_e}{L_e} & \frac{k_v}{L_e}\sin{(\theta_s)} & \frac{k_v V}{L_e}\cos{(\theta_s)} & \frac{V}{V^*_{dc} L_e}\sin{(\theta_s)}\\
         -\eta k_v \cos{(\theta_s)} & -\eta k_v \sin{(\theta_s)} & 2\mu(V^2_0-3V^2)-\frac{\eta P_0}{NV^2} & \eta k_v \xi_2 & -\eta \xi_1/(V^*_{dc})\\
         \frac{\eta k_v}{V} \sin{(\theta_s)} & -\frac{\eta k_v}{V} \cos{(\theta_s)} & \frac{2\eta Q_0}{NV^3}-\frac{\eta k_v \xi_2}{V^2} & \eta k_v \xi_1/V & \eta \xi_2/(V V^*_{dc})\\
         \hline
         -\frac{N V}{C_{dc}V^*_{dc}}\cos{(\theta_s)} & -\frac{N V}{C_{dc}V^*_{dc}} \sin{(\theta_s)} & -N \xi_1/(C_{dc}V^*_{dc}) & N V \xi_2/(C_{dc}V^*_{dc}) & 0\\ 
    \end{array}}\right];\\
    &B_{11}=\left[{\begin{array}{cccc}
         0 & 0 & \frac{\eta}{NV}\cos{(\phi)} & \frac{\eta}{NV^2}\sin{(\phi)}\\
         0 & 0 & \frac{\eta}{NV}\sin{(\phi)} & -\frac{\eta}{NV^2}\cos{(\phi)}\\
    \end{array}}\right]^T;\quad
    B_{22}=\left[{\begin{array}{cc}0 & 0\end{array}}\right],\ \text{where}\  \left[{\begin{array}{c}
         \xi_1\\
         \xi_2\\
    \end{array}}\right] = 
    \left[{\begin{array}{c}
         I_d \cos{(\theta_s)}+I_q \sin{(\theta_s)}\\
         I_d \sin{(\theta_s)}-I_q \cos{(\theta_s)}\\
    \end{array}}\right].
  \end{split}
\end{equation}
\end{figure*}


\bibliographystyle{IEEEtran}
\bibliography{uVOC}

\end{document}